\documentstyle[aps]{revtex}
\twocolumn

	\newtheorem{definition}{Definition}

\newtheorem{Theorem}{Theorem}[section]

\def\Ha{{\cal H}_{aux}}

\def\Cyl{{\rm Cyl}}

\def\be{\begin{equation}}
\def\ee{\end{equation}}
\def\ba{\begin{eqnarray}}
\def\ea{\end{eqnarray}}
\def\Gra{{\rm Gra}}

\def\E{{\cal E}}
\def\A{{\cal A}}
\def\G{{\cal G}}
\def\ag{{{\cal A}/{\cal G}}}

\def\agb{{\overline {{\cal A}/{\cal G}}}}

\def\S{\Sigma}\def\g{\gamma}\def\Gb{{\overline \G}}
\def\C{{\cal C}}
\def\Ab{{\overline \A}}
\def\Gb{{\overline \G}}
\def\B{\agb}
\def\a{\alpha}
\def\b{\beta}
\def\g{\gamma}

\def\Comp{{\mathchoice
{\setbox0=\hbox{$\displaystyle\rm C$}\hbox{\hbox to0pt
{\kern0.4\wd0\vrule height0.9\ht0\hss}\box0}}
{\setbox0=\hbox{$\textstyle\rm C$}\hbox{\hbox to0pt
{\kern0.4\wd0\vrule height0.9\ht0\hss}\box0}}
{\setbox0=\hbox{$\scriptstyle\rm C$}\hbox{\hbox to0pt
{\kern0.4\wd0\vrule height0.9\ht0\hss}\box0}}
{\setbox0=\hbox{$\scriptscriptstyle\rm C$}\hbox{\hbox to0pt
{\kern0.4\wd0\vrule height0.9\ht0\hss}\box0}}}}
\def\Co{{\mathchoice
{\setbox0=\hbox{$\displaystyle\rm C$}\hbox{\hbox to0pt
{\kern0.4\wd0\vrule height0.9\ht0\hss}\box0}}
{\setbox0=\hbox{$\textstyle\rm C$}\hbox{\hbox to0pt
{\kern0.4\wd0\vrule height0.9\ht0\hss}\box0}}
{\setbox0=\hbox{$\scriptstyle\rm C$}\hbox{\hbox to0pt
{\kern0.4\wd0\vrule height0.9\ht0\hss}\box0}}
{\setbox0=\hbox{$\scriptscriptstyle\rm C$}\hbox{\hbox to0pt
{\kern0.4\wd0\vrule height0.9\ht0\hss}\box0}}}}
\def\Rl{{\mathchoice
{\setbox0=\hbox{$\displaystyle\rm R$}\hbox{\hbox to0pt
{\kern0.4\wd0\vrule height0.9\ht0\hss}\box0}}
{\setbox0=\hbox{$\textstyle\rm R$}\hbox{\hbox to0pt
{\kern0.4\wd0\vrule height0.9\ht0\hss}\box0}}
{\setbox0=\hbox{$\scriptstyle\rm R$}\hbox{\hbox to0pt
{\kern0.4\wd0\vrule height0.9\ht0\hss}\box0}}
{\setbox0=\hbox{$\scriptscriptstyle\rm R$}\hbox{\hbox to0pt
{\kern0.4\wd0\vrule height0.9\ht0\hss}\box0}}}}
\newcommand{\tiN}{\raisebox{-6.5pt}{$
\displaystyle \stackrel{\displaystyle N}{\sim} $}}

\def\C{{\cal C}}

\def\o{\overline}

\def\wt{\widetilde}

\def\wh{\widehat}

\def\ag{{\cal A}/{\cal G}}
\def\A{{\cal A}}
\def\ha{{\cal HA}}
\def\hg{{\cal HG}}

\def\H{{\cal H}}
\def\B{{\cal B}}

\def\hab{\overline {\ha}}
\def\O{\Omega}
\def\Ga{\Gamma}

\def\haux{\H_{aux}}
\def\r{\rangle}
\def\l{\langle}
\def\rp{\rangle_{phys}}
\def\rd{\rangle_{diff}}
\def\ra{\r_{aux}}
\def\hp{\H_{phys}}
\def\hd{\H_{diff}}
\def\V{{\cal V}}
\def\vp{\V_{phys}}
\def\vd{\V_{diff}}
\def\bst{{\cal B}^{(\star)}}
\def\baux{{\cal B}_{aux}}
\def\bsa{\bst_{aux}}
\def\bsp{\bst_{phys}}
\def\Ps{\Phi'}
\def\Pgs{\Phi^{\prime}_{[\wt \alpha]}}
\def\ab{\o {\cal A}}
\def\Hva{{\cal H}^{\g, \vec{\pi}}_{aux}}

\def\X{{\cal X}}
\def\Xb{{\overline \X}}

\def\C*{$C^{\star}$}
\def\ge{\geq}

\let\ssection=\section
\renewcommand{\section}{\setcounter{equation}{0}\ssection}

\begin{document}
\cleardoublepage

\title{Quantization of diffeomorphism invariant
theories of connections with local degrees of freedom}

\author{Abhay Ashtekar}
\address{
Center for Gravitational Physics and
Geometry, Physics Department,\\ Penn State University, University Park,
PA 16802-6300, USA}

\author{Jerzy Lewandowski}
\address{Institute of Theoretical Physics,
University of Warsaw,\\ 00-681 Warsaw, Poland}

\author{Donald Marolf}
\address{Department of Physics, The University of
California,\\ Santa Barbara, CA 93106}

\author{Jos\'e Mour\~ao}
\address{Sector de F\'{\i}sica, U.C.E.H.,
Universidade do Algarve,\\ Campus de Gambelas, 8000 Faro, Portugal}

\author{Thomas Thiemann}
\address{Center for Gravitational Physics and
Geometry, Physics Department,\\ Penn State University, University Park,
PA 16802-6300, USA}


\maketitle

\pagenumbering{arabic}

\vfil\eject

\begin{abstract}

Quantization of diffeomorphism invariant theories of connections is
studied. A solutions of the diffeomorphism constraints is
found. The space of solutions is equipped with an inner product that
is shown to satisfy the physical reality conditions. This provides, in
particular, a quantization of the Husain-Kucha\v{r} model. The main
results also pave way to quantization of other diffeomorphism
invariant theories such as general relativity. In the Riemannian case
(i.e., signature ++++), the approach appears to contain all the
necessary ingredients already. In the Lorentzian case, it will have to
combined in an appropriate fashion with a coherent state transform to
incorporate complex connections.
\end{abstract}

\begin{section} {Introduction}
\label{intro}

Keeping with the theme of the special issue, this paper will address
the problem of quantization of a class of diffeomorphism invariant
field theories.

The class can be specified as follows. We will assume that the theory
can be cast in a Hamiltonian form. The configuration variable will be
a connection 1-form $A_a^i$ on a $d$-dimensional (``spatial'')
manifold and takes values in the Lie algebra of a compact, connected
Lie-group. The canonically conjugate momentum, $\tilde{E}^a_i$, will
be a vector field with density weight one (or, equivalently, a $d$-1
form) which takes values in the dual of the Lie algebra. The phase
space $\Gamma$ will thus consist of pairs $(A_a^i,
\tilde{E}^a_i)$ satisfying suitable regularity conditions. Finally, the
gauge invariance will be ensured by the Gauss constraint and the
($d$-dimensional) diffeomorphism invariance, by a vector constraint,
such that the entire system is of first class in Dirac's
terminology. Individual theories in this class may have additional
features such as specific Hamiltonians or additional constraints.  In
the main discussion, however, we will ignore such structures and focus
only of the features listed above which will be common to all theories
in the class.

To make this general setting more concrete, let us list a few
illustrative examples of theories which are included in this class.
The first is the Husain-Kucha\v r model \cite{1} which can be thought
of as general relativity without the Hamiltonian constraint.  Thus, in
this model, we only have the Gauss and the (``spatial'')
diffeomorphism constraints and the Hamiltonian is a linear combination
of them. In this case, we will be able to obtain a complete quantum
theory. A second example is provided by Riemannian (i.e., ++++)
general relativity, cast in a Hamiltonian framework using self-dual
connections. In this case, in addition to the Gauss and the
diffeomorphism constraint, there is also the Hamiltonian constraint
which dictates ``time evolution.'' The results of this paper provide
only a partial solution to the problem of quantization of this model
since the Hamiltonian constraint will not be incorporated. However, as
we will indicate in the last section, the general methods employed
appear to be applicable also to the Hamiltonian constraint and the
issue is currently being investigated.  Next, one can also consider
Lorentzian general relativity in terms of a spin connection and its
conjugate momentum.  Our results will again provide a complete
solution to the Gauss and the diffeomorphism constraints.  (The
Hamiltonian constraint is, however, more difficult to address now.
One possible approach is to pass to self-dual connection
variables\cite{2} using the coherent state transform of Ref. [3].)
Finally, our class allows for Chern-Simons theories whose group is the
inhomogeneous version \cite{8} $IG$ of a compact, connected Lie group
$G$. This class includes Riemannian general relativity in 3 space-time
dimensions.

{}From a mathematical physics perspective, one faces two types of
problems while quantizing such models. First, the underlying
diffeomorphism invariance poses a non-trivial challenge: We have to
face the usual field theoretic difficulties that are associated with
the presence of an infinite number of degrees of freedom but now {\it
without} recourse to a background space-time geometry. In particular,
one must introduce new techniques to single out the quantum
configuration space, construct suitable measures on it to obtain
Hilbert spaces of states and regulate operators of physical
interest. The second set of problems arises because of the presence of
constraints. In particular, even after one has constructed a Hilbert
space and regularized the constraint operators, one is left with the
non-trivial task of {\it solving} the constraints to isolate the
physical states and of introducing an appropriate inner product on
them. This is a significant problem even for systems with only a {\it
finite} number of degrees of freedom since, typically, solutions to
constraints fail to lie in the initial Hilbert space. Thus, physical
states do not even have a natural ``home'' to begin with! In theories
now under consideration, these difficulties become particularly
severe: Diffeomorphism invariance introduces an intrinsic non-locality
and forces one to go beyond the standard techniques of local quantum
field theory.

Our approach to solving these problems is based on two recent
developments.  The first is the introduction of a new functional
calculus on the space of connections modulo gauge transformations
which respects the underlying diffeomorphism invariance (see
Ref. [5-12]). The second is a new strategy for solving quantum
constraints which naturally leads to an appropriate inner product on
the physical states (see Ref.  [6-13]). Together, the two developments
will enable us to complete the general algebraic quantization program
\cite{A2,AT} for the class of systems under consideration. Thus, we
will be able to solve the quantum constraints and introduce the
appropriate Hilbert space structure on the resulting space of
solutions.

The main ideas underlying these developments can be summarized as
follows. Recall first that, in gauge theories, it is natural to use
the space $\ag$ of connections modulo local gauge transformations as
the classical configuration space. In quantum field theories, due to
the presence of an infinite number of degrees of freedom, the quantum
configuration space is typically an enlargement of its classical
counterpart. The enlargement is non-trivial because the measures which
define the scalar product tend to be concentrated on
``distributional'' fields which lie outside the classical
configuration space. In gauge theories, if we require that the Wilson
loop variables --i.e., the traces of holonomies-- should be
well-defined also in the quantum theory, a canonical enlargement
$\agb$ of $\ag$ becomes available \cite{AI}. This space can be thought
of as a limit of the configuration spaces of lattice
gauge theories for all possible ``floating'' (i.e., not necessarily
rectangular) lattices. Geometric structures on configuration spaces of
lattice gauge theories can therefore be used to induce geometric
structures on $\agb$. \cite{AL1,B2,MM,AL2}This enables one to introduce
integral and differential calculus on $\agb$ {\it without} reference
to any background geometry. The calculus can, in turn, be used to
introduce measures, Hilbert spaces of square-integrable functions and
regulated operators on them.

The strategy of solving quantum constraints, on the other hand, is
quite general and not tied to the theories of connections.
\cite{DD,29} For simplicity, consider the case when there is just one
constraint, $C = 0$, on the classical phase space. To quantize the
system, as in the standard Dirac procedure, one first ignores the
constraint and constructs an auxiliary Hilbert space $\Ha$, ensuring
that the set of ``elementary'' real functions on the full phase space
is represented by self-adjoint operators on $\Ha$. Thus, $\Ha$
incorporates the ``kinematic reality conditions''. Since the classical
constraint is a real function on the phase space, one represents it by
a self-adjoint operator $\hat{C}$ on $\Ha$. The solutions are to be
states which are annihilated by $\hat{C}$, or, alternatively, which
are left invariant by the 1-parameter group $U(\lambda) := \exp
i\lambda \hat{C}$ generated by $\hat{C}$. A natural strategy
\cite{AH2,AH3} to obtain solutions, then, is to begin with a suitable
state $\phi$ in $\Ha$ and average it over the group; formally,
$\bar{\phi} := \int d\lambda U(\lambda)\circ|\phi>$ is group
invariant. The problem is that, typically, $\bar{\phi}$ does not
belong to $\Ha$; it is not normalizable. However, it often has a
well-defined action on a dense subset $\Phi$ of $\Ha$ in the sense
that $\bar{\phi}\cdot\ \psi := \int d\lambda <\phi|U(\lambda)\circ |
\psi>$ is well-defined for all $\psi> \in \Phi$.  That is,
$\bar{\phi}$ can be often thought of as an element of the topological
dual of $\Phi$ (if $\Phi$ is equipped with a suitable topology which
is finer than the one induced by $\Ha$). To summarize, group averaging
{\it can} lead to solutions of the quantum constraint but they lie in
a space $\Phi' $ which is larger than $\Ha$ (if, as is typically the
case, zero lies in the continuous part of the spectrum of
$\hat{C}$). Finally, one can introduce an Hermitian inner product on
the space of the solutions simply by setting $<\bar{\phi}_1 |
\bar{\phi}_2> = \bar{\phi}_1\ \cdot \phi_2$. Thus, {\it if} one can
find a dense subspace $\Phi$ in $\Ha$ (and equip it with a suitable
topology) such that the group averaging procedure maps every element
of $\Phi$ to a well-defined element of $\Phi'$, one can extract the
Hilbert space of physical states. One can show that the resulting
physical Hilbert space automatically incorporates the ``reality
conditions'' on physical observables \cite{DD,DM1} {\it even when they
are not known explicitly}.

The purpose of this paper is to use these two developments to obtain
the following results for the class of models under consideration:
\begin{itemize}
\item[1.] We will construct the {\it quantum} configuration space
$\agb$ and select the measure $\mu_0$ on it for which $L^2(\agb,
d\mu_0)$ can serve as the auxiliary Hilbert space $\Ha$, i.e., can be
used to incorporate the kinematical reality conditions of the
classical phase space.
\item[2.] Introduce the diffeomorphism constraints as well-defined
operators on $\Ha$ and demonstrate that there are no anomalies in
the quantum theory.
\item[3.] Construct a dense subspace $\Phi$ of $\Ha$ with the
required properties and obtain a complete set of solutions of the
diffeomorphism constraints in its topological dual $\Phi'$. We will
also characterize the solutions in terms of generalized knots (i.e.,
diffeomorphism invariance classes of certain graphs) and obtain the
Hilbert spaces of physical states by introducing the inner products
which ensure that real physical observables are represented by
self-adjoint operators.
\end{itemize}
\noindent While the main emphasis of the paper is on presenting a
rigorous solution to the diffeomorphism constraint, along the way, we
will summarize a number of additional results which are likely to be
useful more generally. First, we will exhibit an orthonormal basis in
$\Ha$, introduced by Baez \cite{24a} drawing on spin networks
considered by Rovelli and Smolin \cite{24} (see also \cite{23}).
Second, we will present a rigorous transform that maps the states in
the connection representation (i.e., in $\Ha$) to functions on the
loop space. Furthermore, using the orthonormal basis, we will also
present the {\it inverse} transform \cite{23} from the loop
representation \cite{G,RS,S} to the connection representation.
Finally, in the case when $d=3$ and the gauge group is $SU(2)$, using
differential calculus on $\agb$ we will indicate how one can
introduce, on $\Ha$, regulated self-adjoint operators corresponding to
areas of 2-surfaces. The spectra of these operators are discrete and
provide a glimpse into the nature of quantum geometry that underlies
Riemannian quantum general relativity.

The plan of the paper is as follows. Sec.\ref{s3} contains an outline
of the general quantization program. Sec.\ref{class} specifies the
precise class of theories considered and presents in greater detail
models, mentioned above, that are encompassed by our discussion.
Sec.\ref{s4} recalls the structure of the quantum configuration space
$\agb$.  In Sec.\ref{kin}, we construct the auxiliary Hilbert space
$\Ha$ and show that a complete set of real-valued functions on the
classical phase space is indeed promoted to self-adjoint operators on
$\Ha$. We also present the Baez orthonormal basis and discuss the loop
transform and its inverse. The diffeomorphism constraints are
implemented in Sec.\ref{phy} using a series of steps that handle
various technical difficulties.  Sec.\ref{diss} summarizes the main
results and puts them in a broader perspective.

A number of results which clarify and supplement the main discussion
are presented in appendices. Appendix A illustrates some subtleties
associated with the group integration procedure in the case when the
Poisson algebra of constraints is Abelian. Appendix B summarizes the
projective techniques that lie at the heart of the diffeomorphism
invariant functional calculus on $\agb$. Appendix C points out that
the requirement of diffeomorphism invariance has certain technical
consequences that might not have been anticipated easily. Finally,
Appendix D illustrates how one can use the projective techniques to
introduce well-defined operators on $\agb$ which capture geometric
notions such as areas of surfaces and volumes of regions. The
operators can be made self-adjoint on $L^2(\agb , d\mu_0)$ and have
discrete spectra. These results provide a glimpse into the nature of
quantum geometry.

\end{section}

\begin{section} {Quantization outline} \label{s3}

In Ref. [17-18], the Dirac quantization program for constrained
systems was extended to incorporate certain peculiarities of
diffeomorphism invariant theories such as general relativity. In this
section, we will further refine that program using the ``group
averaging'' techniques mentioned in Sec. \ref{intro}. These techniques provide
a
concrete method for constructing solutions to the quantum constraints
{\it and} for introducing an appropriate scalar product on the space
of these solutions.

In the first part of this section, we will spell out the refined
version of the program, and in the second, illustrate the various
steps involved by applying them to three simple examples.

\begin{subsection} {Strategy} \label{s31}

Consider a classical system with first class constraints $ C_i = 0 $
for which the phase space $\Ga$ is a real symplectic manifold. The
proposal is to quantize this system in a series of steps. (The steps
which have been modified from Ref. [17,18] are identified with a
prime.)

\begin{itemize}

\item[Step 1.] Select a subspace $\cal S$ of the vector space of
all smooth, complex-valued functions on $\Ga$ subject to the
following conditions:

\begin{itemize}

\item[a)] $\cal S$ should be large enough so that any sufficiently
regular function on the phase space can be obtained as
(possibly a suitable limit of) a sum of products
of elements in $\cal S$.

\item[b)] $\cal S$ should be closed under Poisson brackets,
i.e. for all functions $F, G$ in $\cal S$, their Poisson bracket
$\{F, G \}$ should also be an element of $\cal S$.

\item[c)] Finally, $\cal S$ should be closed under complex conjugation; i.e.
for all $F$ in $\cal S$, the complex conjugate $F^*$ should be
a function in $\cal S$.

\item[] Each function in $\cal S$ is to be regarded
as an {\it elementary classical variable} which is to have
an {\it unambiguous} quantum analog.

\end{itemize}

\item[Step 2.] Associate with each element $F$ in
$\cal S$ an abstract operator $\wh F$. Construct the
free associative algebra generated by these {\it elementary quantum
operators}. Impose on it the canonical commutation relations,
$[\wh F, \wh G] = i \hbar \wh{ \{ F, G \} }$, and, if necessary,
also a set of (anti-commutation) relations that captures the algebraic
identities satisfied by the elementary classical variables.
Denote the resulting algebra by $\B_{aux}$.

\item[Step 3.] On this algebra, introduce an involution operation
$\star$ by requiring that if two elementary classical variables $F$
and $G$ are related by $F^* = G$, then $\wh F^\star = \wh G$ in
$\baux$. Denote the resulting $\star$-algebra by ${\cal
B}^{(\star)}_{aux}$.

\end{itemize}
(Recall that an involution on $\baux$ is an anti-linear map $\star$ from
$\baux$ to itself satisfying the following three conditions for all
$A$ and $ B$ in $\baux$: i) $(A + \lambda B)^\star = A^\star +
\lambda^* B^\star$, where $\lambda$ is any complex number; ii)
$(AB)^\star = B^\star A^\star$; and iii) $(A^\star)^\star = A$.)

These steps are the same as in Ref. [17,18]. The main idea in the
remaining steps was to use the ``reality conditions'' --i.e., the
requirement that a suitable class of classical observables be
represented by self-adjoint operators-- to determine the inner product
on physical states. This strategy has been successful in a number of
examples \cite{AT}, including a model field theory that mimics several
features of general relativity \cite{TT}. For the class of theories now
under consideration, however, we will refine the remaining steps along
the lines of Ref. [13-16].

While we will retain the idea that the classical reality conditions
should determine the inner product, we will not need to explicitly
display a complete set of classical observables (i.e., functions which
Poisson commute with the constraints).  Instead, we will work with the
complete set of functions (${\cal S}$) on the {\it unconstrained}
phase space, noting that the reality properties of such functions will
determine the reality properties of the observables.  The idea is then
to implement the reality conditions of operators in $\bsa$ on an {\it
auxiliary} Hilbert space $\haux$ from which the physical phase space
$\hp$ will be finally constructed.

\begin{itemize}

\item[Step 4$'$.]  Construct a linear $\star$-representation $R$ of
the abstract algebra $\bsa$ via linear operators on an
auxiliary Hilbert space $\haux$, i.e. such that
\be
R(\wh A^\star) =  R(\wh{A})^\dagger
\ee
for all $\wh {A}$ in $\bst$, where $\dagger$ denotes Hermitian conjugation
with respect to the inner product in $\haux$.

\end{itemize}

We now wish to construct the physical Hilbert space $\hp$, which
will in general {\it not} be a subspace of $\haux$.  We  proceed
as follows.

\begin{itemize}

\item[Step 5$'$a.] Represent the constraints $C_i$ as self-adjoint
operators $\wh C_i$ (or, their exponentiated action, representing the
finite gauge transformations, as unitary operators $\wh U_i$) on
$\haux$.

\end{itemize}

This step provides a quantum form of the constraints that we will use
to define observables and physical states.  We will look for solutions
of the constraints in terms of {\it generalized} eigenvectors of $\wh
C_i$ which will lie in the topological dual $\Phi'$ of some dense
subspace $\Phi \subset \haux$ (see also Ref. [19,27].  Since $\Phi$
and $\Phi'$ will be used to build the physical Hilbert space, we will
consider only physical operators that are well behaved with respect to
$\Phi$.

\begin{itemize}

\item[Step 5$'$b.] Choose a suitable dense subspace $\Phi \subset
\haux$ which is left invariant by the constraints $\hat{C_i}$ and let
$\bsp$ be the $\star$-algebra of operators on $\haux$ which commute
with the constraints $\wh{C}_i$ and such that, for $A \in \bsp$, both
$A$ and $A^\dagger$ are defined on $\Phi$ and map $\Phi$ to itself.
\end{itemize}

Note that the choice of $\Phi$ is subject to two conditions: on the
one hand it should be large enough so that $\bsp$ contains a
``sufficient number'' of physically interesting operators, and, on the
other, it should be small enough so that its topological dual $\Phi'$
is ``sufficiently large'' to serve as a home for physical states.  The
key idea now is to find an appropriate map $\eta: \Phi \rightarrow
\Phi'$ such that $\eta(\phi)$ is a solution to the constraint for all
$\phi \in \Phi$. (Note that the natural class of maps from $\Phi$ to
$\Phi'$ is {\it anti}-linear (c.f., the adjoint map)).

\begin{itemize}

\item[Step 5$'$c.]  Find an anti-linear map $\eta$ from $\Phi$ to the
topological dual $\Ps$ that satisfies:

\begin{itemize}

\item [(i)] For every $\phi_1 \in \Phi$, $\eta(\phi_1)$ is a solution
of the constraints; i.e.,

\be
0 = \bigl(\wh{C}_i (\eta \phi_1)\bigr)[\phi_2] := (\eta \phi_1) [
\wh{C}_i \phi_2]
\ee
for any $\phi_2 \in \Phi$.
Here, the square brackets denote the natural action of $\Ps$ on
$\Phi$.

\item[(ii)] $\eta$ is  real and positive in the sense that, for all
$\phi_1,\phi_2 \in \Phi$,
\ba \label{rp}
({\eta \phi_1})[\phi_2] &=& ((\eta \phi_2)[\phi_1])^* {\qquad {\rm and}
\qquad } \nonumber\\
(\eta \phi_1)[\phi_1] &\ge& 0.
\ea

\item[(iii)] $\eta$ commutes with the action of any $A \in \bsp$ in the
sense that \be
\label{stareq}
(\eta \phi_1)[A \phi_2] = ((\eta A^\dagger \phi_1))[\phi_2]
\ee

for all $\phi_1,\phi_2 \in \Phi$.
\end{itemize}

\end{itemize}
(The appearance of the adjoint on the r.h.{s.} of (\ref{stareq})
corresponds to the anti-linearity of $\eta$.)

\begin{itemize}

\item[Step 5$'$d.] The vectors $\eta \phi$ span a space $\vp$ of
solutions of the constraints. We introduce an inner product on $\vp$
through
\be
\label{pip}
\l \eta \phi_1, \eta \phi_2 \rp = (\eta \phi_2) [ \phi_1]
\ee
The requirement (\ref{rp}) guarantees that this inner product is well
defined and that it is Hermitian and positive definite.
Thus, the completion of $\vp$ with respect to (\ref{pip}) is a
`physical' Hilbert space $\hp$.
\end{itemize}
\noindent (Note that the positions of $\phi_1$ and $\phi_2$ must be
opposite on the two sides of (\ref{pip}) due to the anti-linear nature
of $\eta$.)

At this point, the reader may fear that this list of conditions on
$\eta$ will never be met in practice.  That the new step 5$'$ may
actually simplify the quantization program follows from the
observation of \cite{AH2,AH3} (and \cite{DD,DM1} for the case when the
Poisson algebra of constraints is Abelian) that a natural candidate
for such a map exists.

Let us indicate, heuristically, how this can come about.  Assume that
the exponentiated form of all constraints $\wh C_i$ defines the
unitary action ($\wh U$) of a group (of gauge transformations) $K$ on
$\haux$. Then, a natural candidate for the map $\eta$ is provided by
the ``group averaging procedure''. Set

\be
\label{gav}
\eta |\phi\r := (\int_K dk \wh U(k) |\phi\r)^\dagger \
= \int_K dk \l \phi| \wh U^{-1}(k)\  ,
\ee
where $dk$ denotes a bi-invariant measure on $K$ (or, rather, on the
orbit through $|\phi\r$), and ignore, for the moment, the issue
convergence of the integral in (\ref{gav}). Then, it is easy to check
that $\eta$ satisfies properties (i)-(iii) in 5$'$c.  Finally, the
expression (\ref{pip}) of the scalar product reduces to:
\be
\label{gavip}
\l \eta \phi_1, \eta \phi_2 \rp =  \int_K dk \l \phi_2 |U^{-1}(k)|
\phi_1 \ra.
\ee
Thus, it is intuitively clear that the requirements of step 5 can be
met in a large class of examples.

Let us return to the general program. The last step is to represent
physical operators on $\vp$. This is straightforward because the
framework provided by step 5 guarantees that $\hp$ carries an (anti)
$\star$-representation (see below) of $\bsp$ as follows:

\begin{itemize}

\item[Step 6$'$.]  Operators in $ A \in \bsp$ have a natural action
(induced by duality) on $\Ps$ that leaves $\vp$ invariant. Use this
fact to induce densely defined operators $A_{phys}$ on $\hp$ through
\be
\label{physop}
A_{phys}\ (\eta \phi) = \eta ( A \phi).
\ee

\end{itemize}

This leads to an {\it anti}- $\star$-representation of $\bsp$ in the
sense that the map (\ref{physop}) from $\bsp$ to the operators on
$\hp$ is and anti-linear $\star$-homomorphism.  Thus, {\it the reality
properties of the physical operators $\bsp$ on $\haux$ descend to the
physical Hilbert space}.

We conclude this subsection with two remarks.  Suppose, first, that
for some $A \in \bsp$ we have $A = A^\dagger$ on $\haux$.  If the
operators $(A \pm i)^{-1}$ are both defined on ${\Phi}$ and preserve
${\Phi}$, then the range of $A_{phys}\pm i$ contains $\vp$ and is
dense in $\hp$. It then follows that $A_{phys}$ is essentially
self-adjoint \cite{RS1} on $\hp$.  The second remark has to do with
our restriction to strong observables, i.e., observables which commute
with constraints. On physical grounds, on the other hand, one should
deal with more general, weak observables. It is often the case that
every weak observable of the system is weakly equivalent to a strong
observable. In these cases, our restriction does not lead to a loss of
generality. In more general cases, on the other hand, an extension of
this procedure to encompass weak observables is needed.
\end{subsection}

\begin{subsection} {Examples}
\label{ex}

We will now present three examples to illustrate how the group
averaging procedure can be carried out in practice. (Parameterized
Newtonian particles and some other examples are treated in Ref. [16]
and appendix \ref{supsel} contains general comments on the case of
Abelian constraints.)  The non-trivial application of this procedure
to diffeomorphism invariant theories will be given in Sec. \ref{phy}.

{\vskip .3 cm } {\it Example A} {\vskip .3 cm }

As a first test case, let us consider a non-relativistic particle in
three dimensions subject to the classical constraint $p_z = 0$, so
that the associated gauge transformations are just translations in the
$z$-direction.  Since the interesting classical functions can be built
from $x,y,z,p_z,p_y,p_z$, we let these six functions span the
classical subspace ${\cal S}$ of step 1 and construct the algebra
$\bsa$ of step 3.  We choose the auxiliary Hilbert space to be ${\cal
H}_{aux} = L^2(\Rl^3,dxdydz)$ and let $\wh x, \wh y, \wh z$ act by
multiplication and $\wh p_x,\wh p_y, \wh p_z$ act by ($-i$ times)
differentiation so that all six operators are self-adjoint.

Clearly, our physical states will be associated with {\it generalized}
eigenstates of $\wh p_z$.  We wish to view such states as
distributions that act on some dense subspace $\Phi \subset \haux$.
With our choice of operators, it is natural to take $\Phi$ to be the
space of smooth functions with compact support.  Note that the Fourier
transform $\tilde f_0$ of any such function $f_0$ is smooth. Hence,
for any $g_0 \in \Phi$, the distribution $\eta({g_0}) := \tilde g^*_0
\delta(p_z)$ has well defined action on any $f_0 \in \Phi$:
\be
\label{e39}
\eta({g_0})[f_0] = \int_{\Rl^3} \tilde g_0^*(p) \delta(p_z) \tilde f_0(p)
dp_xdp_ydp_z \ ,
\ee
where, as before, $*$ denotes complex conjugation.  Note that this
action may be constructed by averaging over the translation group
through
\be
\label{iav}
\eta(g_0) [f_0] = \int_{\Rl^3} dxdydz \int_{\Rl} dz' g^*_0(x,y,z+z')
f_0(x,y,z)\ .
\ee

We now let $\vp$ be the linear space spanned by such $\eta(g_0)$.
This space is annihilated by $\wh p_z$ (under the dual action of $\wh
p_z$ on $\Ps$) and will become a dense subspace of the physical
Hilbert space $\hp$.

For $f,g$ in $\vp$, let $f_0$ be an element of $\Phi$ that maps to $f$
under $\eta$. Then, our prescription (\ref{e39})) yields the following
physical inner product: $\langle f,g \rangle_{phys} = g[f_0]$, where
$f_0$ may be any smooth function $f_0(x,y,z)$
of compact support for which $f(x,y) = \int dz f^*_0(x,y,z)$; i.e.,
$\eta (f_0) = f$.  Thus, the physical inner product is just $\int
f^*(x,y)g(x,y) dxdy$.  It is Hermitian, positive definite, and
independent of the choice of $f_0$.  The resulting ${\cal H}_{phys}$
is just what one would expect on intuitive grounds and, since the
observables $\wh x,\wh y,\wh p_x,\wh p_y$ act on $\vp$ by
multiplication and ($-i$ times) differentiation, they define
self-adjoint operators on $\hp$ and the reality conditions are
satisfied in the usual way.

Finally, note that there is a freedom to scale the map $\eta$ by a
constant: for real positive $a$, the use of $\eta_a = a \eta$ would
simply re-scale the physical inner product by an overall factor and
lead to an equivalent physical Hilbert space.  This freedom can be
traced back to the fact that the Haar measure ($dz'$ in \ref{iav}) on a
non-compact group is unique only up to a multiplicative factor.

{\vskip .3 cm }
\goodbreak
{\it Example B}
{\vskip .3 cm }

Our second example (also treated in Ref. [14,16]) will be the
massive free relativistic particle in four-dimensional Minkowski
space.  Recall that this system may be classically described by a
phase space $\Rl^8$ with coordinates $x^\mu,p_\nu$ for $\mu,\nu \in
\{0,1,2,3\}$.  It is subject to the constraint $p^2 +m^2 = 0$ and has
an associated set of gauge transformations which may be loosely
interpreted as `time reparametrizations.'  Again, these classical
functions define the space ${\cal S}$ of step 1 and the algebra $\bsa$
of step 3.  Thus, we represent them by self-adjoint
quantum operators $\wh x^\mu, \wh p_\nu$ which act on the auxiliary
Hilbert space $L^2(\Rl^4,d^4x)$ by multiplication and ($-i$ times)
differentiation.  We will concentrate on the dense space $\Phi$ of
smooth functions with compact support, so that elements $f_0$ of
$\Phi$ have smooth Fourier transforms $\tilde f_0$.

Let us attempt to apply the group averaging technique and define
$\eta(f_0)$ (for $g_0 \in \Phi$) such that, for any $g_0 \in \Phi$,
\be
\label{fp}
\eta({f_0})[g_0]
= \int_{\Rl^4} d^4x \int_{\Rl} d\lambda [\exp(i\lambda
\wh C) f_0^*](x)g_0(x)\ ,
\ee
where $\wh C = \wh p^2 + m^2$. By spectral analysis, we know that the
Fourier transform $\wt{f}$ of $f=\eta (f_0)$ is just $\tilde f^*_0
\delta (p^2 + m^2)$ so that (\ref {fp}) does in fact define an
element of $\Phi'$.

The span of such $f$ defines the linear space $\vp$.  Now, for any
$f,g \in \vp$, choose some $f_0$ such that $f = \eta(f_0)$ and define
$\l f, g \rp = g[f_0]$.  Note that the inner product (\ref{fp}) is is
just the integral of $\tilde{f}^*_0(p)\tilde{g}_0(p)$ over the mass
shell.  This inner product is manifestly positive definite, Hermitian,
and independent of the choice of $f_0$ and $g_0$.  Thus, the resulting
$\hp$ is the usual Hilbert space associated with the free relativistic
particle, except that it contains both the `positive and negative
frequency parts' as orthogonal subspaces.  While none of the operators
$\wh x^\mu,\wh p_\nu$ are observables, they can be used to construct
observables on $\haux$ for which the induced operators on $\hp$ are
the familiar Newton-Wigner operators (see Ref. [16, 29]).  Again, any
of the maps $\eta_a = a \eta$ may be used in this construction.

{\vskip .3 cm }
\goodbreak
{\it Example C}
{\vskip .3 cm }

Finally, we consider what we will call the massive free relativistic
particle on a globally hyperbolic, curved four dimensional space-time
${\cal M}$ with metric $g_{\mu \nu}$.  We will allow an arbitrary
space-time for which the wave operator $\nabla_\mu
\nabla^\mu$ is essentially self-adjoint when acting on the Hilbert
space $L^2({\cal M},dv)$, where $dv$ is the space-time
volume element.

We take the classical phase space to be $\Gamma = T^*M$, but subject
our system to the constraint $ g^{\mu \nu}(x) p_\mu p_\nu +m^2 = 0$.
Here, $p_\mu$ is the four-momentum and this constraint generates an
associated group of gauge symmetries. We choose smooth functions on
${\cal M}$ and $V^\mu p_\mu$ for complete vector fields $V^\mu$ on
${\cal M}$ to generate the subspace ${\cal S}$ and the algebra $\bsa$.
It is then natural to choose $\haux$ to be $L^2({\cal M},dv)$ and to
represent real functions on ${\cal M}$ by self-adjoint operators that
act by multiplication.  Similarly, real complete vector fields $V^\mu$
are represented by the self-adjoint differential operators $(-i) V^\mu
\partial_\mu -{i \over 2} {\rm div} (V)$, where ${\rm div}(V)$ denotes
the divergence of $v$ with respect to the space-time metric; ${\cal
L}_V dv = {\rm div}(V) dv$.  The constraint is promoted to the unique
self-adjoint extension $\wh C$ of the wave operator on $L^2({\cal M},
dv)$. (The freedom to add a multiple of the scalar curvature of
$g_{\mu\nu}$ does not affect the discussion that follows.)

It is again natural to take $\Phi$ to be the space of smooth functions
on ${\cal M}$ with compact support.  We then define the map $\eta:
\Phi \rightarrow \Ps$ by
\be
(\eta f_0)(x) = \bigl(
\int_{\Rl} d \lambda e^{i \lambda \wh C} f_0^* \bigr) (x)
\ee
and take $\vp$ to be its image.  Here we appeal to Gel'fand spectral
theory \cite{22} to show that the resulting generalized eigenstates
lie in the topological dual $\Phi'$ of $\Phi$.  As before, the natural
concept is in fact the family of maps $\eta_a = a \eta$ for $a \in
\Rl^+$.  The physical Hilbert space $\hp$ is the completion of $\vp$
in the inner product $\l f , g\rp = g[f_0]$ where $f_0$ satisfies $f =
\eta_1 (f_0)$.  This inner product is independent of the particular
choice of $f_0$, is Hermitian and positive definite, and self adjoint
operators $A$ on $\haux$ which preserve $\Phi$ and commute with $\wh
C$ induce symmetric, densely defined operators $A_{phys}$ on $\hp$.

The construction of $\hp$ may come as a surprise to some readers as it
seems to violate the accepted idea that there is no well-defined
notion of a single relativistic quantum particle in a non-stationary
space-time.  The `resolution' is that the quantum theory defined above
does not exhibit the properties that one would require for it to
describe a `physical' free particle.  In particular, it contains no
notion of a conserved probability associated with Cauchy surfaces, as
our particle appears to `scatter backwards in time' when it encounters
a lump of space-time curvature. (Re-collapsing cosmologies \cite{DM1}
illustrate a similar effect).  In addition, this framework cannot be
used as the one-particle Hilbert space to build a relativistic field
theory.  Recall that an essential element in the construction of a
quantum field from a one particle Hilbert space is that the inner
product on the Hilbert space be compatible with the symplectic
structure on the space of classical solutions (which is given by the
Klein-Gordon inner product).  That this is not the case for our inner
product may be seen from the fact that it contains no notion of a
conservation law associated with Cauchy surfaces.

\end{subsection}

\end{section}

\begin{section} {The class of theories}
\label{class}

In this section, we spell out in some detail the class of
theories to be considered and discuss various features which
will be used in subsequent sections.  The section is divided into
three parts.  We present the general framework in the first,
some illustrative examples of theories satisfying
our assumptions in the second, and in the third, a set of functions on the
phase
spaces of these theories which will serve as elementary variables
in the quantization program.

\begin{subsection} {General framework}
\label{class1}

Let suppose that the underlying ``space-time'' $M$ is a $d+1$
dimensional manifold with topology $M=\Rl\times\Sigma$ where $\Sigma$
is an orientable, real analytic, $d$ dimensional manifold.  We wish to
consider field theories on $M$ which admit a Hamiltonian formulation
with following features:\\
a) The phase space consists of canonical pairs $(A_a^i,\tilde{E}^a_i)$
where $A_a^i$ is a connection 1-form on $\Sigma$ taking values in
the Lie algebra of a compact, connected gauge group $G$, and
$\tilde{E}^a_i$, its conjugate momentum, is a vector density of weight
one on $\Sigma$ which takes values in the dual of the Lie algebra of
$G$. The fundamental Poisson brackets will be:
\be \label{2.1}
\{A_a^i(t,x),\tilde{E}^b_j(t,y)\}=\delta (x,y)\delta^i_j\delta_a^b \;.
\ee
b) The theory is a constrained dynamical system subject to (at least)
the following two constraints:
\ba \label{2.2}
G_i &:=& (\partial_a \tilde{E}^a+[A_a,\tilde{E}^a])_i = 0\
\mbox{ and}\\
V_a &:=& \mbox{tr}[F_{ab}\tilde{E}^b] = 0 \
\ea
where $F$ is the curvature of $A$.  The first of these will be
referred to as the {\it Gauss constraint} and the second as the {\it
vector} or the {\it diffeomorphism constraint}.  A given theory in the
class may well have other constraints.

It is easy to check that the canonical transformations generated by
the Gauss constraint correspond to local gauge transformations
associated with $G$ while those associated with (a suitable
combination of the Gauss and) the vector constraint correspond to
diffeomorphisms of $\Sigma$. The constraint algebra formed by these
two constraints is of first class.  The action of these theories will
have the general form:
\ba \label{2.3}
S=\frac{1}{c^2}\int_\Rl dt\int_\Sigma
d^dx & &(\mbox{tr}[\dot{A}_a\tilde{E}^a]-
[-\mbox{tr} [\Lambda G]+\nonumber\\
& &N^a V_a +\mbox{other terms}]) \ ,
\ea
where $c$ is a coupling constant, $\Lambda^i,N^a$ are associated
Lagrange multipliers and ``other terms" could contain additional
constraints. (For simplicity, we have left out possible boundary
terms.)  We will assume that the full system of constraints is of
first class and that the Hamiltonian is (weakly) invariant under the
canonical transformations generated by {\em all} constraints.

In the following sections, for most part, we will focus just on the
Gauss and the vector constraints.
\end{subsection}

\begin{subsection}{Example theories}
\label{s22}

In this section, we will provide several examples to illustrate
the type of theories that are encompassed by our analysis.
\\

A) {\it The Husain-Kucha\v{r} model}

\noindent This is perhaps the simplest non-trivial example. Here, the
gauge group $G$ is $SU(2)$ and the manifold $\Sigma$ is 3-dimensional.
As mentioned in the Introduction, it has no further constraints and
the Hamiltonian is a linear combination of the two constraints.
Somewhat surprisingly, the model does arise from a manifestly
covariant, 4-dimensional action \cite{1}. Although it is not of direct
physical significance, this model is interesting from a mathematical
physics perspective because it has all the features of general
relativity except the Hamiltonian constraint.
\\

B) {\it Riemannian general relativity}

\noindent A second model is provided by 4-dimensional general relativity
with metrics of signature (++++). Again, at least at first sight, this
model is not of direct physical interest. However, since it contains
all the conceptual non-trivialities of Lorentzian general relativity,
it provides an excellent arena to test various quantization
strategies. Furthermore, there are some indications that, if one were
to solve this model completely, one may be able to pass to the quantum
theory of Lorentzian general relativity by a ``generalized Wick
rotation'' which would map suitably regular functions of the Euclidean
self-dual connections to holomorphic functions of the Lorentzian
self-dual connections.

Since this model is not discussed in the literature, we will write
down the basic equations governing it. We will, however be brief since
the Lorentzian counterpart of this case has been analyzed in detail in
Ref.[2,17]. The key idea here is to use a Palatini-type of action,
however with {\it self-dual} connections. Thus, we begin with:
\be \label{2.4}
S({}^4\!A, e) =\int_M d^4x (e) e^a_I e^b_J\ ({}^{4}\!F)_{ab}^{IJ}
\ee
where the $a,b,c$ are the four-dimensional tensor indices,
$I,J,K=1,..,4$ are the ``internal'' $SO(4)$ indices, $e^a_I$ is a
tetrad (for a positive definite metric), $e$ its determinant,
${}^4\!A_a^{IJ}$, a {\it self-dual} connection and
${}^4\!F_{ab}^{IJ}$, its curvature. Although we are using self-dual
connections, the variation of this action provides precisely the
vacuum Einstein's equations.

For simplicity, let us assume that the 3-manifold $\Sigma$ is compact.
(The asymptotically flat case requires more care but can be treated in
an analogous fashion \cite{2,A2}.) Then, if we perform a 3+1
decomposition, let $t^a$ be the ``time-evolution'' vector field, and
use a suitable basis in the 3-dimensional self-dual sub-algebra of the
$SO(4)$ Lie-algebra, we can cast the action in the form:
\ba\label{2.9}
S =\int_\Rl dt\int_\Sigma d^3x & &[\tilde{E}^a_i
{\cal L}_t A_a^i - \nonumber\\
& &\{-A_t^i G_i+N^a V_a +{1 \over 2}\tiN C\}] \;.
\ea
Here indices $a,b,...$ refer to the tangent space of $\Sigma$ and
$i,j,...$ to the self-dual ($SU(2)$) Lie algebra;
$A_t:=t^a\;{}^{4}\!A_a$, $N^a$ and $\tiN$ are Lagrange multipliers;
and, $(A_a^i, \tilde{E}^a_i)$ are the canonical variables.  Thus,
symplectic structure is given by \be
\label{2.11} \{A_a^i(x),\tilde{E}^b_j(y)\}=\delta_a^b\delta^i_j\;
\delta^{(3)}(x,y) \;.
\ee
The variation of the action with respect to the Lagrange multipliers
yields, as usual, the first class constraints of Riemannian general
relativity:
\label{2.10} \ba
G_i&:=& {\cal D}_a\tilde{E}^a_i \equiv
\partial_a\tilde{E}^a_i +\epsilon_{ijk}
A_a^j\tilde{E}^a_k  = 0 \ ,\nonumber\\
V_a &:=& F_{ab}^i \tilde{E}^b_i = 0 \ , {\rm and}\nonumber\\
C&:=& F_{abi}\tilde{E}^a_j\tilde{E}^b_k\epsilon^{ijk} = 0\ .
\ea
These are, respectively, the Gauss, the vector and the scalar
constraint. Thus, in the Hamiltonian form, the theory is similar to
the Husain-Kucha\v{r} model except for the presence of the
additional scalar constraint.

How do we make contact with the more familiar Hamiltonian form of the
theory in terms of metrics and extrinsic curvatures? The two are
related simply by a canonical transformation. Regard $\tilde{E}^a_i$ as a
triad on $\Sigma$ with density weight one and denote by $\Gamma_a^i$
the spin-connection defined by it. Define $K_a^i$ via:
$K_a^i=\Gamma_a^i- A_a^i$. Then, $(A_a^i, \tilde{E}^a_i) \mapsto
(\tilde{E}^a_i, K_a^i)$ is a canonical transformation. $\tilde{E}^a_i$
determines the 3-metric $q_{ab}$ on $\Sigma$ via $\tilde{E}^a_i
\tilde{E}^{bi} = q q^{ab}$, and $K_a^i$ determines the extrinsic curvature
$K_a{}^b$ via $\sqrt{q}K_a{}^b = K_a^i \tilde{E}^b_i$, where $q$ is
the determinant of $q_{ab}$. Note, however, that, while the
constraints (\ref{2.10}) are all low order polynomials in terms of the
connection variables, they become non-polynomial in terms of the
metric variables. Hence, if one uses the metric formulation, it is
much more difficult to promote them to well-defined operators on an
auxiliary Hilbert space.
\\

C) {\it Lorentzian general relativity in the spin connection
formulation}

\noindent In the Lorentzian signature, self-dual connections are complex.
Therefore, the formulation of the Lorentzian theory in terms of
self-dual connections \cite{2,A2} falls outside the scope of this
paper. However, as in the Euclidean case, one can consider the real
fields $(\tilde{E}^a_i, K_a^i)$ as a canonical pair. By a contact
transformation, one can replace the triad $\tilde{E}^a_i$ by the
spin-connection $\Gamma_a^i$ and $K_a^i$ by the momentum
$\tilde{P}^a_i$ conjugate to $\Gamma_a^i$. In the new canonical pair,
the configuration variable is a $SU(2)$ connection whence the
framework falls in the class of theories considered here. One can show
that the Gauss and the vector constraints retain their form; $A_a^i$
and $\tilde{E}^a_i$ in (\ref{2.11}) are simply replaced by $\Gamma_a^i$
and $\tilde{P}^a_i$ respectively. Therefore, in this formulation, the
theory belongs to the class under consideration.

Unfortunately, however, the remaining, scalar constraint seems
unmanageable in terms of $\Gamma_a^i$ and $\tilde{P}^a_i$. Hence, this
formulation is not directly useful beyond the Gauss and the vector
constraints \cite{5}. As mentioned in the Introduction, to handle the
Hamiltonian constraint, one would have to use a different strategy,
e.g., the one involving a coherent state transform \cite{6} and pass
to the (Lorentzian) self-dual representation.

D) {\it Chern-Simons theories}

\noindent Let $G$ may be any compact, connected Lie group. Then, one can
construct a natural ``inhomogeneous version'' $IG$ of $G$. As a
manifold, $IG$ is isomorphic to the cotangent bundle over $G$ and, as
a group, it is a semi-direct product of $G$ with an Abelian group
which has the same dimension as $G$. If $G$ is chosen to be the
rotation group, $SO(3)$, then $IG$ is the Euclidean group in three
dimensions.  (For details, see Ref. [4,31].) Let us now set the
dimension $d$ of $\Sigma$ to be $2$ and consider the Chern-Simons
theory based on $IG$. (If $G$ is chosen to be $SU(2)$, this theory is
equivalent to 3-dimensional Riemannian general relativity.)  It is
straightforward to check that all our assumptions from
Sec. \ref{class1} are satisfied.

We can also consider a more sophisticated enlargement $I_\Lambda G$ of
$G$ which is parametrized by a real number $\Lambda$ (see
Ref. [31]). In the case when $G$ is $SU(2)$, the Chern-Simons
theory based on $I_\Lambda G$ is the same as Riemannian general
relativity with a cosmological constant. (Curiously, the theory that
results from $G=SU(2)$ and $\Lambda$ negative is also isomorphic, in
an appropriate sense, with the Lorentzian, 3-dimensional gravity with
a {\it positive} cosmological constant.)  All these theories also fall
in the class under consideration.  Note however that, in general, the
Chern-Simons theories based on compact gauge groups $G$ --rather than
$IG$ or $I_\Lambda G$-- fall outside this class since these theories do
not have canonical variables of the required type.
\end{subsection}

\begin{subsection} {An (over)complete set of gauge invariant functions}
\label{ls}

In this section, for simplicity of presentation, we will focus on the
case $d=3$ and $G=SU(2)$. Generalizations to higher dimensions and
other compact, connected groups is, however, straightforward.  For
simplicity, we will solve the Gauss constraint classically. (See,
however, the first part of Sec. \ref{diss}.)  Therefore, it is natural
to regard the space $\ag$ of (sufficiently well-behaved) connections
on $\Sigma$ modulo (sufficiently regular) gauge transformations as the
effective configuration space. Phase space is then the cotangent
bundle on $\ag$. Our aim is to single out a convenient set of
functions on this phase space which can be used as ``elementary
classical variables'' in the quantization program of Sec. \ref{s3}.

Wilson loop functions are the obvious candidates for configuration
variables. These will be associated with piecewise analytic loops on
$\Sigma$, i.e., with piecewise analytic maps $\alpha: S^1 \mapsto
\Sigma$. (Thus, the loops do not have a preferred parameterization,
although in the intermediate stages of calculations, it is often
convenient to choose one.) The Wilson loop variables $T_\alpha (A)$
are given by:
\be T_\alpha (A) := \mbox{tr}\ h_\alpha [A] \equiv
\mbox{tr} {\cal P} \exp\int_\alpha A \ ,
\ee
where the trace is taken in the fundamental representation.  As
defined, these are functions on the space of connections. However,
being gauge invariant, they project down naturally to $\ag$. The
momentum observables, $T_S$ are associated with piecewise analytic
strips $S$, i.e., ribbons which are foliated by a 1-parameter family
of loops. For technical reasons, it is convenient to begin with
piecewise analytic embeddings $S: (1, 1) \times S^1 \mapsto \Sigma$
and use them to generate more general strips. Set
\ba \label{2.14}
T_S(A) &:= &\int_{S} dS^{ab} \eta_{abc} T^c_{\alpha_\tau} (
\sigma, \tau)\ , \quad \mbox{ where} \nonumber\\
T^c_{\alpha_\tau} (\sigma, \tau) &:=&
\mbox{tr}(h_{\alpha_\tau}(\sigma, \tau)[A]\tilde{E}^c(\sigma, \tau)),
\ea
$\sigma, \tau$ are coordinates on $S$ (with $\tau$ labeling the loops
within $S$ and $\sigma$ running along each loop $\alpha_\tau$),
$\eta_{abc}$ denotes the Levi-Civita tensor density on $\Sigma$, and,
as before $h_{\alpha_\tau}$ denotes the holonomy along the loop
$\alpha_\tau$.  Again, the functions $T_S$ are gauge invariant and
hence well-defined on the cotangent bundle over $\ag$. They are called
``momentum variables'' because they are linear in $\tilde{E}^a_i$.

Properties of these variables are discussed in some detail in
Ref. [17]. Here we recall only the main features.  First they
constitute a complete set in the sense that their gradients span the
cotangent space almost everywhere on the phase space over
$\ag$. However, they are not all independent.  Properties of the trace
operation in the fundamental representation of $SU(2)$ induce
relations between them. These algebraic relations have to be
incorporated in the quantum theory.  It is interesting that the
Poisson brackets can be expressed in terms of simple geometric
operations between loops and strips.  We will ilustarate this by
writing out one of these Poisson brackets which will be needed in the
subsequent sections:
\be \label{2.16}
\{T_\alpha,T_S\}=\sum_i \mbox{sgn}_i(S,\alpha)
[T_{S\circ_i\alpha}-T_{S\circ_i\alpha^{-1}}]
\ee
where the sum is over transverse intersections $i$ between the the
loop $\alpha$ and the strip $S$, ${\rm sgn}_i(S,\alpha)$ takes values
$0, \pm 1$ depending on the orientation of the tangent vector of
$\alpha$ and the tangent plane of $S$ at the $i$-th intersection point
and $S\circ_i\alpha$ is a loop obtained by composing the loop in the
strip $S$ passing through the intersection point with the loop
$\alpha$.  (Note that the same geometric point in $\Sigma$ may
feature in more than one intersection $i$.) Thus, in particular, the
Poisson bracket vanishes unless the loop $\alpha$ intersects the strip
$S$.

The Poisson bracket between two strip functionals also vanishes unless
the two strips intersect. If they do, the bracket is given by a sum
of slightly generalized strip functionals. The generalization consists
only of admitting certain strip maps $S: (1,1)\times S^1 \mapsto
\Sigma$ which is not necessarily embeddings, and integrating in
(\ref{2.14}) over a suitable sub-manifold $I$ without boundary,
$I\subset (0,1)\times S^1$, such that for every loop $\alpha_\tau$,
$\alpha_\tau\cap S(I)$ is a closed loop. The Poisson bracket between
these more general strips closes. We did not simply begin with these
more general strips because, in quantum theory, it is easier to
begin with the embedded strips and let them generate more general
ones.

In Sec. \ref{kin}, we will use these loop and strip functionals as
elementary classical variables to construct the auxiliary Hilbert
space.

\end{subsection}
\end{section}

\begin{section}  {Quantum configuration space} \label{s4}

To complete the first four steps in the quantization program, it is
convenient to proceed in two stages. First, one focuses on just the
configuration variables $T_\alpha$ and constructs representations of
the corresponding ``holonomy algebra.'' This naturally leads to
the notion of a quantum configuration space. By introducing suitable
geometric structures on this space, one can then represent the momentum
operators corresponding to $T_S$.

We will begin, in this section, by isolating the quantum configuration
space. In the second part, we will present three convenient
characterizations of this space. A number of constructions used in the
subsequent sections depend on these characterizations. In the third
part, we introduce elements of calculus on this space which will lead
to the definition of the momentum operators in Sec. \ref{kin}.

\begin{subsection} {$\agb$ a completion of $\ag$} \label{s41}

In the classical theory, $\ag$ serves as the gauge invariant
configuration space for the class of theories under consideration. We
will now show that, in the passage to quantum theory, one is led to
enlarge this space \cite{AI}.  Recall that an enlargement also
occurs in, for example, scalar quantum field theory \cite{GJ,R}.

Let us begin by constructing the Abelian algebra of configuration
operators.  This algebra is, of course, generated by finite linear
combinations of functions $T_\a$ on $\ag$ with complex
coefficients. By construction, it is closed under the operation of
taking complex conjugation. Thus, it is a $^\star$-subalgebra of the
algebra of complex-valued, continuous bounded functions on $\ag$. It
separates the points of $\ag$ in the sense that, if $[A_1]\neq [A_2]$
(i.e., if the gauge equivalence classes of $A_1$ and $A_2$ in ${\cal
A}$ do not coincide), there exists a loop $\a$ such that: $T_\a(A_1)
\neq T_\a(A_2)$. Thus, as indicated in Sec. \ref{ls}, the set of
configuration variables is sufficiently large. This algebra is called
the {\it holonomy algebra} and denoted by $\ha$. To obtain a greater
degree of control, it is convenient to introduce on it a norm and
convert it into a $C^\star$ algebra.

Let us therefore set
\be
\parallel f \parallel = \sup_{[A] \in \ag} \mid f( [A]) \mid
\label{3.5}
\ee
and complete $\ha$ with respect to this norm we obtain a commutative
$C^\star$-algebra $\o \ha$. (This algebra is equipped with identity,
given by $T_\emptyset$, where $\emptyset$ is the trivial, i.e., point
loop.) We are now in a position to apply the powerful representation
theory of $C^\star$-algebras.

The first key result we will use is the Gel'fand-Naimark theorem, that
every $C^\star$ algebra with identity is isomorphic to the
$C^\star$-algebra of all continuous bounded functions on a compact
Hausdorff space called the {\it spectrum} of the algebra. The spectrum
can be constructed directly form the algebra: it is the set of all
$\star$-homomorphisms from the given $C^\star$-algebra to the
$\star$-algebra of complex numbers.  We will denote the spectrum of
$\o \ha$ by $\agb$. It is easy to show that $\ag$ is densely embedded
in $\agb$; thus, $\agb$ can be regarded as a completion of $\ag$.

Recall that, since $\o\ha$ is the $C^\star$-algebra of configuration
variables, our primary objective here is to construct its
representations.  Now, a key simplification occurs because one has a
great deal of control on the representation theory. Let $\rho :
\o \ha \rightarrow B({\cal H})$ denote a  cyclic representation of
$\o \ha$ by bounded operators on some Hilbert space ${\cal H}$. Let
$\Gamma$ be the ``vacuum expectation value functional'':
\be
\Ga (f) = \langle \rho(f) \O, \O \rangle \ ,       \label{3.7}
\ee
where $\Omega$ is a cyclic vector and $f$ any element of $\o \ha$.
Clearly, $\Gamma$ is a positive linear functional on $\o\ha$. Since $\o
\ha$ is isomorphic with the $C^\star$-algebra of continuous functions
on $\agb$, $\Gamma$ can be regarded as a positive linear functional
also on $C^0(\agb )$. Now, since $\agb$ is compact, the Riesz
representation theorem ensures that there is a a unique regular Borel
measure $\mu$ on $\agb$ such that
\be
\Ga (f) = \int_{\agb} d \mu( {[A]} ) \wt f({[A]} )   \ ,  \label{3.8}
\ee
where $\tilde{f} \in C^0(\agb)$ corresponds to $f$ in $\o \ha$.  This
immediately implies that any cyclic representation of $\hab$ is
unitarily equivalent to a ``connection representation'' given by
\ba
\hab & \rightarrow& B (L^2 (\agb , \mu))  \nonumber \\
(\rho(f) \psi)( {[A]}) & =& \wt f( {[A]})
\psi( {[A]})  \   ,
\label{3.9}
\ea
where the measure $\mu$ is defined through (\ref{3.8}). Therefore the
set of regular measures on $\agb$ is one-to-one correspondence with
the set of cyclic representations of $\o \ha$.

To summarize, in {\it any} cyclic representation of $\o \ha$, quantum
states can be thought of as (square-integrable) functions on $\agb$
(for some choice of measure. Recall that cyclic representations are
the basic ``building blocks'' of general representations.)  Hence,
$\agb$ can be identified with the {\it quantum configuration
space}. The enlargement from $\ag$ to $\agb$ is non-trivial because,
typically, $\ag$ is contained in a set of zero measure \cite{MM}.

We will conclude this subsection with a general remark. In the
construction of the quantum configuration space, we have avoided the
use of the non-gauge invariant affine structure of the space ${\cal
A}$ of connections and worked instead directly on $\ag$.  (For earlier
works in the same spirit, see \cite{AM}.)  This is in contrast with
with the gauge fixing strategy that is sometimes adopted in
constructive quantum field theory \cite{GJ,R} which then faces global
problems associated with Gribov ambiguities.

\end{subsection}

\begin{subsection} {Characterizations of $\agb$}
\label{s42}

Since $\agb$ is the domain space of quantum states, it is important to
understand its structure. In this subsection, therefore, we will
present three characterizations of this space, each illuminating its
structure from a different perspective.

Denote by ${\cal L}_{x_0} \S$ the space of continuous piecewise
analytic loops on $\S$ based at an arbitrarily chosen but fixed point
$x_0$. Two loops $\a, \b$ are said to be holonomically equivalent if
for every $A \in \A$ we have \be H(\a, A) = H(\b, A) \ . \label{3.12}
\ee The corresponding equivalence classes are called {\it hoops}. For
notational simplicity we will use lower case greek letters to denote
these classes as well.  The set of all hoops forms a group called the
hoop group which is denoted by $\hg_{x_0}$. A smooth connection $A \in
\A$ defines a homomorphism from $\hg_{x_0}$ to $SU(2)$, which is
smooth in a certain sense \cite{Ba}
\be H(. , A) \ : \ \hg_{x_0} \
\rightarrow \ SU(2) \ .  \label{3.13}
\ee
We can now present the first characterization: $\agb$ is naturally
isomorphic to the set of {\it all} homomorphisms from $\hg_{x_0}$ to
$SU(2)$ modulo conjugation \cite {AL1}. (The conjugation serves only
to eliminate the freedom to perform gauge transformations at the base
point. Note that the homomorphism here need not even be continuous.)
This result makes it possible to show further that $\agb$ is a
limit of configuration spaces of gauge theories living in
arbitrary lattices for which the space of connections modulo
gauge transformations coincides with finite products of copies of
$SU(2)$ modulo conjugation \cite{MM,AMM,AL2,AL3}.

The second characterization is in terms of these limits. To introduce
it, let us begin with the notion of independent hoops\cite{AL1}. Hoops
$\{ \b_1, ..., \b_n \}$ will be said to be independent if loop
representatives exist such that each contains an open segment that is
traced exactly once and which intersects other representatives at most
at a finite number of points.  Let now $S_n(\b_1, ..., \b_n)$ denote
the subgroup of $\hg_{x_0}$ generated by a set of independent hoops
$\{\b_1, ..., \b_n \}$. The space $H(S_n)$ of all homomorphisms
(modulo conjugation) from $S_n$ to $SU(2)$ is homeomorphic to
$SU(2)^n/Ad $, which in turn can be thought of as the configuration
space of the ``floating'' (i.e., non-rectangular) lattice formed by
$\{\b_1, ..., \b_n \}$. Now, if we consider a larger subgroup $S_m
\supset S_n$ of the hoop group, we have a natural projection map
$p_{S_n S_m}$, where
\ba p_{S_nS_m} \ : \ H(S_m) \ &\rightarrow& H(S_n) \nonumber \\
p_{S_n S_m}(h) &=& h|_{S_n} \ .
\label{ccs}
\ea
In the lattice picture, the projection is obtained simply by
restricting the configurations on the larger lattice to the smaller
lattice.

The family $(H(S_n), p_{S_n S_m})$ is called a {\it projective family}
labeled by the subgroups $S_n$ of the hoop group (see appendix
\ref{proj}). Since the theory for a larger lattice contains more
information, it is desirable to consider larger and larger lattices,
i.e., bigger and bigger subgroups of the hoop group. Unfortunately the
projective family itself does not have a ``largest element'' from
which one can project to any other. However, such an element can in
fact be obtained by a standard procedure called the ``projective
limit.''. Now, given the space $\agb$, we have a surjective
projection
$p_{S_n}$ to $H(S_n)$ for {\it any} subgroup $S_n$ of the hoop group:
\ba
p_{S_n} \ : \ \agb  \ &\rightarrow&  \ SU(2)^n / Ad
\nonumber \\
 {[A]} \ & \mapsto & \ ( [{A} (\b_1)], ...,  [{A}(\b_n)])
\label{3.15} \ea
where the brackets $[,]$ on the right hand side denote conjugacy
classes. This suggests $\agb$ may be the projective limit of the
family $(H(S_n), p_{S_nS_m})$. Detailed considerations show that this
is indeed the case \cite{MM}.

This characterization of $\agb$ as a limit of finite dimensional
spaces allows the introduction of integral calculus
\cite{AL1,B2,L,AMM,AL3} on $\agb$ using integration theory
on finite dimensional spaces. Roughly, measures on lattice
configuration spaces $H(S_n)$ which are compatible with the
projections $P_{S_nS_m}$ from larger lattices to the smaller ones
induce measures on the projective limit $\agb$. In particular, this
strategy was first used in \cite{AL1} to construct a natural, faithful,
diffeomorphism invariant measure $\mu_0$ on $\agb$ from the induced
Haar measures on the configuration spaces $H(S_n)$ of lattice
theories. More precisely, $\mu_0$ is defined by:
\be
p_{S_n \star} \mu_0 = p_{Ad^\star} \mu_H \otimes ... \otimes
\mu_H             \ .
\label{3.16}
\ee
where, $\mu_H$ denotes the Haar measure on $SU(2)$, $p_{Ad}$ denotes
the quotient map
\ba
p_{Ad} \ : \ SU(2) \times .&.&. \times SU(2) \ \rightarrow \nonumber\\
SU(2) \times ... \times &SU(2)&  / Ad   \ ,
   \label{3.17}
\ea
and $f_\star \mu$ denotes the push-forward of the measure $\mu$ with
respect to the map $f$.

This description uses hoops as the set of ``probes'' for the
generalized connections.  A related approach, developed by
Baez,\cite{B2} relies on the (gauge dependent) probes defined by
analytic edges.  This strategy provides a third characterization of
$\agb$, again as a projective limit, but of a projective family
labeled by graphs rather than hoops. It is this characterization that
is best suited for developing differential calculus
\cite{L,AL2}. Since it is used in the subsequent sections, we will
discuss it in greater detail.

Let us begin with the set $\cal E$ of all oriented, unparametrized,
embedded, analytic intervals (edges) in $\Sigma$. We introduce the
space $\Ab$ of (generalized) connections on $\Sigma$ as the space of
all maps ${ A}\ :\ {\cal E}\ \rightarrow SU(2)$, such that
\be
\label{10}
 A(e^{-1}) =  [A(e)]^{-1} , \  {\rm and} \
{ A}(e_2\circ e_1)\ =\ { A}(e_2){ A} ( e_1)
\ee
whenever two edges $e_2,e_1\in{\cal E}$ meet to form an edge.  Here,
$e_2 \circ e_1$ denotes the standard path product and $e^{-1}$ denotes
$e$ with opposite orientation.  The group $\o\G$ of (generalized)
gauge transformations acting on $\Ab$ is the space of {\it all} maps $
g:\Sigma \rightarrow SU(2)$ or equivalently the Cartesian product group
\be
\label{11} \Gb\ := \times_{x\in\S} \ SU(2) \ .
\ee
A gauge transformation $ g \in \Gb$ acts on $ A \in \Ab$ through
\be
[{g}({A})](e_{p_2,p_1}) =
({g}_{p_2})^{-1} {A}(e_{p_2,p_1}) {g}_{p_1} \ ,
\ee
where $e_{p_2,p_1}$ is an edge from $p_1 \in \Sigma$ to $p_2 \in
\Sigma$ and ${g}_{p_i}$ is the group element assigned to $p_i$ by
${g} $.  The group $\Gb$ equipped with the product topology is a
compact topological group.  Note also that $\Ab$ is a closed subset of
the Cartesian product of all $\A_e$,
\be
\Ab\subset \times_{e\in\E} \ \A_e \ ,
\ee
where the space $\A_e$ of all maps from the one point set $\{e\}$ to $
SU(2)$ is homeomorphic to $SU(2)$.  $\Ab$ is then compact in the topology
induced from this product.

The space $\Ab$ (and also $\Gb$) can also be regarded as the
projective limit of a family labeled by graphs in $\Sigma$ in which
each member is homeomorphic to a finite product of copies of
$SU(2)$. \cite{MM,AL3}  Let us now briefly recall this construction as
it underlies the introduction of calculus on $\agb$.

\begin{definition}  A graph on $\Sigma$ is a finite subset
$\g\subset\E$ such that $(i)$ two different edges, $e_1, e_2 \ : \ e_1
\neq e_2 $ and $ e_1 \neq e_2^{-1}$, of $\g$ meet, if at all, only at
one or both ends and $(ii)$ if $e \in \g$ then $e^{-1} \in \g$.
\end{definition}

The set of all graphs in $\Sigma$ will be denoted by $\Gra(\S)$.  In
$\Gra(\S)$ there is a natural relation of partial order $\ge$,
\be
\g'\ \ge\ \g
\ee
 whenever every edge of $\g$ is a path product of
edges associated with $\g'$.  Furthermore, for any two graphs
$\gamma_1$ and $\g_2$, there exists a $\g$ such that $\g \geq \g_1$
and $\g \ge \g_2$, so that $(\Gra(\S), \ge)$ is a directed set.

Given a graph $\gamma$, let ${\cal A}_{\gamma}$ be the associated
space of assignments (${\cal A}_{\g} = \{A_{\g} | A_{\g}: \g
\rightarrow SU(2) \}$) of group elements to edges of $\gamma$, satisfying
$A_\g(e^{-1}) = A_\g(e)^{-1}\mbox{ and }A_\g(e_1\circ e_2)=A_\g(e_1)
A_\g(e_2)$, and let $p_{\gamma} : \o{{\cal A}}
\rightarrow {\cal A}_{\gamma}$ be the projection which restricts
${A} \in \o{\cal A}$ to $\g$. Notice that $p_\g$ is a surjective
map.  For every ordered pair of graphs, $\g'\ge\g$, there is a
naturally defined map
\be
\label{0.15} p_{\g\g'}\ :\ \A_{\g'}\
\rightarrow \ \A_\g, \ \ {\rm such}\ \ {\rm that}\ \ p_{\g}\
= \ p_{\g\g'} \circ p_{\g'}   \ .
\ee
With the same graph $\g$, we also associate a group $\G_\g$ defined by
\be
\G_\g\ :=\ \{g_{\g} |g_{\g} : V_{\gamma} \rightarrow SU(2)\} \ ,
\ee
where $V_{\g}$ is the set of {\it vertices} of $\g$; that is, the set
$V_\g$ of points of $\S$ lying at the ends of edges of $\g$.  There is a
natural projection $\Gb\ \rightarrow\ \G_\g$ which will also be
denoted by $p_\g$ and is again given by restriction (from $\S$ to
$V_\g$). As before, for $\g'\ge\g$, $p_{\g}$ factors into $p_{\g}\ = \
p_{\g\g'} \circ p_{\g'}$ to define
\be
\label{0.17}
p_{\g\g'}\ :\ \G_{\g'}\ \rightarrow \G_\g \ .
\ee
Note that the group $\G_\g$ acts naturally on $\A_\g$ and that this
action is equivariant with respect to the action of $\Gb$ on $\Ab$ and
the projection $p_\g$. Hence, each of the maps $p_{\g\g'}$ projects to
new maps also denoted by
\be
\label{0.18}
p_{\g\g'} :\ \A_{\g'}/{\cal G}_{\g'}\
\rightarrow \ \A_{\g}/{\cal G}_{\g} \ .
\ee

We collect the spaces and projections defined above into a (triple)
projective family $(\A_\g, \G_\g, \A_\g/\G_\g,
p_{\g\g'})$.  It is not hard to see that $\Ab$ and
$\Gb$ as introduced above are just the projective limits of the first
two families.  Finally, the quotient of compact projective limits is
the projective limit of the compact quotients, \cite{AL3}
\be
\label{B}
\Ab/\Gb\ =  \agb \ .
\ee
This concludes our third characterization of $\agb$. (Note that the
projections $p_{\g\g'}$ in (\ref{0.15}), (\ref{0.17}) and (\ref{0.18})
are different from each other and that the same symbol $p_{\g\g'}$ is
used only for notational simplicity; the meaning should be clear from
the context.)

Using again the normalized Haar measure on $SU(2)$, the construction
(\ref{3.16},\ref{3.17})  may be repeated for this projective family
\cite{B2}. This leads to a  natural (``Haar'') measure ${\mu'}_0$
defined on $\ab$ via
\be \label{mu}
{\mu'}_0\ =\ \{{\mu}_{\g}=\mu_H\otimes...\otimes \mu_H\}.
\ee
Under the natural projection map to $\agb$, the push forward of this
measure yields $\mu_0$ of (\ref{3.16}).

\end{subsection}

\begin{subsection} {Differential calculus on $\agb$} \label{s43}

We now recall from Ref. [11] some elements of calculus on $\agb$
defined using calculus on finite dimensional spaces and the
representation of $\agb$ as a projective limit.  This framework will
allow us, in the next section, to represent $\wh{T}_S$ as operators on
$L^2(\agb, d\mu_0)$.

Although our primary interest is $\agb$, it will be convenient to
introduce geometric structures on $\Ab$. Vector fields and other
operators that are invariant under the action of $\Gb$ on $\Ab$ will
descend to $\agb = \Ab / \Gb$ and provide us with differential
geometry on the quotient.

Let us begin by introducing the space of $C^n$ {\it cylindrical
functions} on $\Ab$ (for details, see appendix \ref{proj}):
\be
\Cyl^n(\Ab)\ = \bigcup_{\g\in\Gra(\S)}(p_{\g})^\star
C^n(\A_\g)  \
\label{e427}
\ee
where $p_\gamma^\star f = f \circ p_\gamma$ is the pull-back to $\Ab$
of the $C^n$ function $f$ on the manifold $\A_\g$. The sub-space of
$\Gb$-invariant functions in $\Cyl^n(\Ab)$ constitutes the space
$\Cyl^n(\agb)$ of $C^n$ cylindrical functions on $\agb$.  Although
any one element of $\Cyl^n(\agb)$ knows only about the restriction of
the $A$ to a graph $\gamma$, since we allow all possible graphs, the
space $\Cyl^n(\Ab)$ is in fact quite large. In the application of the
quantization program, $\Cyl^\infty(\agb)$ will serve as the space
$\Phi$, i.e., the analog of the space of $C^\infty$ functions of
compact support used in the examples in Sec. II.B.

Let us now consider vector fields. These can be regarded as
derivations of the algebra $\Cyl^\infty(\Ab)$, i.e.
\ba
X \ : \ \Cyl^\infty(\Ab) \ &\rightarrow& \ \Cyl^\infty(\Ab)  \\
X(fg) &=& X(f)g + fX(g) \ .     \label{leir}
\ea
A natural way to construct these vector fields is via consistent
families of vector fields $(X_\g)$ on $\A_\g$. This correspondence is
given by the natural measure ${\mu'}_0$ on $\ab$ and
\be
\int_{\A_\g} \o g_\g X_\g (f_\g) d\mu_\g^H  = \int_{\Ab}
\o g X(f) d\mu'_0  \ ,
\label{vvff}
\ee
for all $f_\g, g_\g \in C^1(\A_\g)$, where $f = p_\g^\star f_\g$ and
$g = p_\g^\star g_\g$.  The family $(X_\g)$ is ($\mu'_0$-) consistent
in the sense that for all $\g' \geq \g$, and for all $f_\g, g_\g \in
C^1(\A_\g)$,
\be
\int_{\A_{\g'}} \o {p^\star_{\g \g'} g_\g} X_{\g'}
(p^\star_{\g \g'} f_\g) d\mu_{\g'}^H = \int_{\A_{\g}} \o { g_\g}
X_{\g} (f_\g) d\mu_{\g}^H  \ .
\label{muco}
\ee
The cylindrical vector fields take a particularly simple form if there
exists a $\g_0$ such that
\be\label{strong}
(p_{\g \g'})_\star X_{\g'} = X_\g
\ee
for all $\g' \geq \g \geq \g_0$. These vector fields were introduced
and studied in detail in Ref.[11].  They will play an important role
in the next section for the representation of $\wh T_S$ as operators.

More general cylindrical operators
\be
B \ : \ \Cyl^\infty(\Ab) \ \rightarrow \ \Cyl^\infty(\Ab)
\ee
can be associated with families $(B_\g)$ of operators acting on
$C^\infty(\A_\g)$ and satisfying the same consistency conditions as
vector fields in (\ref{muco}) Examples of such operators are
Laplacians \cite{6,AL2} on $\Ab$ and the geometric operators discussed
in Appendix D.

\end{subsection}

\end{section}

\begin{section} {Quantum Kinematics}
\label{s5}
\label{kin}

We are now ready to apply the algebraic quantization of program of
Sec. \ref{s3} to the class of theories under consideration. In this section,
we will complete the first four steps in the program. We begin by
introducing the auxiliary Hilbert space $\Ha$ which incorporates the
reality conditions on the loop-strip functions and then analyze some
of its structure.

\begin{subsection} {Auxiliary Hilbert space and reality conditions}

Let us use the vector space generated by finite linear combinations
(with constant coefficients) of the loop and strip functionals of
Sec. IIIC as the space ${\cal S}$ of elementary classical variables
and denote by $\B_{aux}^{(\star)}$ the resulting $\star$-algebra. Our
job now is to find a $\star$-representation of this algebra by
operators on a Hilbert space $\Ha$.

Let us choose for $\Ha$ the space $L^2(\agb, d\mu_0)$, where $\mu_0$
is the faithful, diffeomorphism invariant measure on $\agb$ induced by
the Haar measure on the gauge group.  The discussion of Sec. \ref{s41}
tells us that the configuration operators $\hat{T}_\alpha$ should
act by multiplication:
\be
(\hat{T}_\alpha\circ \psi)([A]): = T_\alpha([A]) \psi([A])
\ee
for all $\psi \in L^2(\agb, d\mu_0)$.  By construction, these operators
are (bounded and) self-adjoint; the reality conditions on the
configuration variables are thus incorporated. Note that this would
have been the case for any choice of measure; it is not essential to
choose $\mu_0$ at this stage.

The condition that $\hat{T}_S$ be represented by self-adjoint
operators, on the other hand, does restrict the measure significantly.
Since $T_S$ is linear in momentum, one would expect it to be
represented by the Lie derivative along a vector field on $\agb$. This
expectation is essentially correct. The detailed definition of
$\hat{T}_S$ is, however, somewhat complicated.

Let us begin by introducing a simpler operator from which $\hat{T}_S$
will be constructed.  Consider an analytic loop $\alpha$. we can think
of it as a graph with just one edge. Fix a point $p$ on $\alpha$ and a
$d-2$-dimensional subspace $W$ of the tangent space at $p$. (Recall
that the underlying manifold $\Sigma$ is $d$ dimensional.) Then, given
a graph $\gamma \ge \alpha$, and a function $F_\gamma$ on $\A_\gamma$,
we wish to define the action of a vector field $X_{\alpha,W}$ on
$F_\gamma$. The key idea is to exploit the fact that, if $\gamma$ has
$n$ edges, $(e_1, ..., e_n)$, then $\A_\gamma$ is isomorphic with
$(SU(2))^n$ and can be coordinatized by $n$ group valued coordinates
$(g_1, ..., g_n)$. Using this fact, we set:
\ba \label{5.3.2}
X_{\alpha,W}& &\circ F_\gamma
:=\mbox{tr}(h_\alpha\tau_j)k^{ij} \nonumber\\
& &\sum_{e\in\gamma}[k^-(e)X^{L}_{e,i}+k^+(e)X^{R}_{e,i}]\
\circ F_\gamma
\ea
where
\[ k^\pm(e):= \nonumber\\
\left\{ \begin{array}{ll}
0 & \mbox{ if }e^\pm\not=p\\
\frac{1}{4}[\mbox{sgn}(\dot{e}^\pm,\dot{\alpha}^+,W)
+\mbox{sgn}(\dot{e}^\pm,\dot{\alpha}^-,W)]& \mbox{if} {e^\pm = p}
\end{array} \right.\]
Here, $h_\alpha$ is the (generalized) holonomy function on $\A_\gamma$
associated with the loop $\alpha$, $\tau_i$ are the Pauli matrices,
$k^{ij}$, the metric in the Lie algebra of $SU(2)$, $X^R_{e,i}$ and
$X^L_{e,i}$ are the right and the left invariant vector fields on the
copy of the group associated with the edge $e$ which point in the
$i$-th direction at the identity of the group, $e^\pm$ refers to the
two ends of the edges, sgn$(\dot{e}^\pm, \dot\alpha^\pm, W)$ is
$0, \pm 1$ depending on the relative orientation of the vectors
involved and the subspace $W$, and $\alpha^+$ (respectively, $\alpha^-$)
is the outgoing (incoming) segment of $\alpha$ at $p$.
 While the definition of this vector
field seems complicated at first, it is in fact straightforward to
calculate its action on functions on $\A_\gamma$. In particular, what
counts is only the dependence of the function $F_\gamma$ on the
group elements corresponding to the edges which pass through $p$ for
which the orientation factor is non-zero.

For each $\gamma\ge \alpha$, we now have a vector field on
$\A_\gamma$.  One can check that these vector fields satisfy the
compatibility conditions (\ref{strong})
and thus provides a vector field $(X_\g)$ on $\ab$ which we will again
denote by $X_{\alpha, W}$. The definition then immediately implies that
this vector field is invariant under $\o {\cal G}$. Hence it has a
well-defined action on the space ${\rm Cyl}^1(\agb)$ on $\agb$ of
differential cylindrical functions on $\agb$ and a well defined
divergence with respect to $\mu_0$. \cite{AL2} A direct calculation
shows that
\be
{\rm div}X_{\alpha, W}\ =\ 0.
\ee

We are now ready to define the strip operators. Given a strip $S$
which is {\it analytically} embedded in $\Sigma$, let us set
\be \label{5.3.5}
\hat{T}_S:=-i\hbar\sum_{x\in S} X_{\alpha_x,W_x}
\ee
where $W_x$ is any $(d-2)$ plane through $x$ which is transversal to
the loop $\alpha_x$ in the strip passing through $x$ and tangent to
the strip.  Although there is an uncountably infinite number of loops
involved in this definition $(\ref{5.3.5})$, the action of $\hat{T}_S$
is nonetheless well-defined on cylindrical functions since, in this
action, only a finite number of terms give non-zero contributions.
The simplest cylindrical functions are the traces of holonomies.
On these, the action of $\hat{T}_S$ reduces simply to:
\ba
(&\hat{T}_S&\circ T_\beta) ([A]) =\nonumber \\
 &-i& \hbar \sum_i {\rm sgn}_i (S, \beta)
[T_{S\circ_i\beta} [A] - T_{S\circ_i\beta^{-1}}[A]\ ,
\ea
where we have used the same notation as in (\ref{2.16}).  This is
action that one would have expected on the basis of the Poisson
bracket \ref{2.16}, so that the commutators between $\hat{T}_\beta$
and $\hat{T}_S$ are the required ones.  Finally, using the fact that
each vector field $X_{\alpha, W}$ is divergence-free, one can
show that $\hat{T}_S$ is essentially self-adjoint.
Thus, the representation of these elementary operators does incorporate
all the reality conditions.

We will conclude with two remarks.\\

1. Our strip operators have been directly defined only for
analytically embedded strips. Since more general strip functionals
were generated by Poisson brackets of the analytically embedded ones,
the corresponding operators are obtained by taking commutators between
the ``basic'' strip operators.
2. In the above discussion, we first set $\Ha = L^2(\agb, d\mu_0)$,
introduced loop and strip operators on it, and argued that the
resulting representation of ${\cal B}_{aux}$ satisfies the reality
conditions. There is in fact a stronger result.  One can begin with
cylindrical functions on $\agb$ and define $\hat{T}_\alpha$ and
$\hat{T}_S$ as above. Then, $\mu_0$ is the only non-trivial measure on
$\agb$ for which the reality conditions can be satisfied. (The
qualification ``non-trivial'' is necessary because, as was pointed out
in Sec. III, the loop-strip variables are complete everywhere except
at the flat connections with trivial holonomies and one can introduce
another measure which is concentrated just at that point of $\agb$
which will also incorporate the reality conditions.) Thus, the overall
situation is similar to that in ordinary quantum mechanics where the
the Lebesgue measure is uniquely picked out by the reality conditions
once we specify the standard representation, $-i\hbar
\vec{\nabla}$ of the momentum operator.\\

\end{subsection}

\begin{subsection} {Spin networks and the (inverse) loop transform}
\label{s52}

In this subsection, we recall \cite{24a} that $\Ha$ admits a
convenient basis and point out the relation between the connection and
the loop \cite{G,RS} representations.

Let us begin with the notion of ``spin-networks'' as formulated by
Baez \cite{24a} (see also Ref. [21, 22]).

The geometrical object called {\em spin-network} is a triple
$(\gamma,\vec{\pi},\vec{c})$ consisting of

$(i)$ a graph $\gamma$,

$(ii)$ a labeling $\vec{\pi}:=(\pi_1,..,\pi_n)$ of  edges $e_1,...,e_n$ of
that graph  $\gamma$ with irreducible representations  $\pi_i$  of $G$,

$(iii)$ a labeling $\vec{c}=(c_1,...,c_m)$ of the vertices $v_1,...,
v_m$ of $\g$ with {\it contractors} $c_j$  (see below).

Each contractor $c_j$ is an intertwining operator from the tensor
product of the representations corresponding to the incoming edges at
a vertex $v_j$ to the tensor product of the representations labeling
the outgoing edges.  Because the group $G$ is compact, the vector
space of all possible contractors $c_j$ associated with a given vector
$\vec{\pi}$ and vertex $v_j$ is finite dimensional.

To $(i-iii)$ we add a forth `non-degeneracy' condition,

$(iv)$ for every edge $e$ the representation $\pi_e$ is non-trivial and
$\g$ is a `minimal' graph in the sense that if another graph
$\g'$ occupies the same set of points in $\Sigma$, then each
edge of $\g'$ is contained in an edge of $\g$.  (Equivalently,
$\g'$ can always be built by subdividing the edges of $\g$, but $\g$ cannot
be so built from $\g'$.)

A {\em spin-network state} is simply a $C^\infty$ cylindrical function
on $\agb$ (a $\Gb$ invariant function on $\ab$) constructed from a
spin-network,
\be \label{5.2.2}
T_{\gamma,\vec{\pi},\vec{c}}[A]:=\mbox{tr}[\otimes_{i=1}^n\pi_i
(h_{e_i}(A))\cdot \otimes_{j=1}^m c_j].
\ee
for all $A \in \Ab$, where, as before, $h_{e_i}(A)= A(e_i)$ is an
element of $G$ associated with an edge $e_i$ and `$\cdot$' stands for
contracting, at each vertex $v_j$ of $\g$, the upper indices of the
matrices corresponding to all the incoming edges and the lower indices
of the matrices assigned to all the outgoing edges with all the
indices of $c_j$.

Using the spin-network states it is easy to construct an orthonormal
basis in $\Ha$. To begin, given a pair $\g, \vec{\pi}$, consider the
vector space ${\cal H}^{\g, \vec{\pi}}$ spanned by the spin-network
states $T_{\g,\vec{\pi}, \vec{c}}$ given by all the possible
contractors $\vec{c}$ associated with $\g, \vec{\pi}$ as above.  Note,
that
\be\label{decomposition}
\Ha\ =\ \bigoplus_{\g, \vec{\pi}} {\cal H}^{\g, \vec{\pi}}_{aux}
\ee
where $\g, \vec{\pi}$ ranges over all the pairs of minimal graphs and
labelings by irreducible, non-trivial representations, the sum is
orthogonal and the spaces $\Hva$ are finite dimensional.  Thus, we
need only choose an orthonormal basis in each $\Hva$.  An explicit
construction is given in Ref. [20,22].

We now turn to loop transforms. This discussion will be brief because
it is not used in the rest in the rest of the paper.  Given any
measure $\mu$ on $\agb$ we can perform the integrals
\be \label{5.2.1}
\chi(\alpha_1,..,\alpha_r):=\int_\agb d\mu_0(A)
T_{\alpha_1}(A)...T_{\alpha_r}(A)
\ee
to obtain a function of multi-loops. In the case when $G=SU(n)$,
Mandelstam identities enable us to express finite products of traces
of holonomies in terms of sums of products involving $r$ or less
traces where $r$ is the rank of the group.  Hence, in the loop
representation, we have to deal only with functions of $r$ or less
loops.  On the other hand, by the Riesz-Markov theorem, any positive
linear functional on $C^0(\agb)$ that satisfies the conditions induced
by the Mandelstam identities is the loop transform $\chi$ of a regular
measure supported on $\agb$.  Thus, there is a one to one
correspondence between between regular measures $\mu$ and their
characteristic functions $\chi$.  This result is analogous to the
Bochner theorem that is used in the framework of constructive quantum
field theory \cite{GJ}. In fact, the loop transform can be thought of
as a precise analog of the Fourier transform for a quantum field
theory with a linear quantum configuration space.

We will now indicate how one can explicitly recover the finite joint
distributions of the measure $\mu$ from its characteristic
functional. (Details will appear elsewhere \cite{23}.)  This
reconstruction of the measure can be regarded as the inverse loop
transform. Given a measure $\mu$, choose an orthonormal basis of
spin-network states $T_{\gamma,\vec{\pi},\vec{c_I}}$ and define the
associated spin-network characteristic function to be the analog of
(\ref{5.2.1}), namely
\be \label{5.2.4}
\chi(\gamma,\vec{\pi},\vec{c_I}):=<T_{\gamma,\vec{\pi},\vec{c_I}}> \;.\ee
We will say that the characteristic functional is absolutely summable
if and only if, for any finitely generated graph $\gamma$, the series
\be \label{5.2.5}
\sum_{\vec{\pi}}\sum_{\vec{c_I}=\vec{c_I}(\vec{\pi})}|\chi(\gamma,\vec{\pi},
\vec{c_I})|<\infty
\ee
is absolutely convergent. We can now state the theorem \cite{23} in
question
\begin{Theorem}
Let the loop transform of a measure be such that the characteristic
functional is absolutely summable. Then the associated family of compatible
measures on $\A_\gamma$ is given by:
\ba \label{5.2.6}
d\mu_\gamma(g_1,..,g_n)=\sum_{\vec{\pi}}\sum_{\vec{c_I}}
&T^\star_{\vec{\pi} ,\vec{c_I}}(g_1,..,g_n) \times \nonumber\\
&\chi (\gamma,\vec{\pi},\vec{c_I})
d\mu_H (g_1, ..., g_n)\;.
\ea
\end{Theorem}

This is a precise analogue of the inverse Fourier transform in the
linear case.

\end{subsection}

\end{section}

\begin{section}{The Hilbert Space of Diffeomorphism Invariant States}
\label{phs}
\label{phy}
Our discussion in sections 4 and 5 has served to introduce and study
the auxiliary Hilbert space $\haux = L^2(\agb,d\mu_0)$.  As this space
carries a $\star$-representation of the algebra (\ref{2.14}) defined
by the loop and strip operators ($\hat{T}_\alpha$ and $\hat{T}_S$), we
have implemented steps 1-4 of the refined algebraic quantization
program (see section \ref{s31}).  In the present section, we will
complete the remaining steps (5 and 6) and construct the Hilbert space
of diffeomorphism invariant states.  For simplicity, we assume
throughout this section that the underlying manifold $\Sigma$ is
$\Rl^3$ (although the results on $\Rl^n$ are identical).

A key step in our construction will involve an appropriate averaging
of spin-network states over the diffeomorphism group. This averaging
procedure was considered, independently, by John Baez\cite{jbaa} as a
tool for constructing a rich variety of diffeomorphism invariant
measures on $\agb$.

\begin{subsection} {Formulation of the diffeomorphism constraint}

Recall that the diffeomorphism constraint is given by:

\be
V_a(x) :=  tr[F_{ab}(x) \wt E^b(x)] = 0\ .
\label{difc}
\ee
Let us considered the smeared version of this constraint,
\be
\label{Ndiff}
V_N:= \int_{\Rl^3} N^a(x) V_a(x) d^3x = 0\ ,
\ee

where $N^a$ are complete analytic vector fields on $\Sigma$. (We
require analyticity because the edges of our graphs are assumed to be
analytic. See Sec. \ref{s4} and \ref{kin}.) Denote by $\varphi_t$ the
1-parameter family of diffeomorphisms generated by $N^a$ on
$\Sigma$. Now, as shown in Appendix \ref{unex}, $V_N$ has a natural
action on the space of smooth functions on $\ag$ which can be used to
define a 1-parameter family $U(t)$ of unitary operators on $\Ha$,
providing us a faithful, unitary representation of the group
$\varphi_t$. On spin network states, the action of the operator
$U_\varphi$ corresponding to $\varphi$ is given by:
\be
U_\varphi\circ(T_{\alpha, \vec \pi,\vec c}) = T_{\alpha,\vec \pi,
\vec c} \circ \varphi =  T_{\varphi \alpha, \varphi \vec \pi,
\varphi\vec c }\ ,
\ee
where $\varphi \alpha$ is the image of the graph $\alpha$ under the
analytic diffeomorphism and $\varphi \vec \pi$ and $\varphi \vec c$
are the corresponding vector of representations and contraction
associated with the new graph $\varphi \alpha$.

Thus, as needed in the group averaging procedure, each
constraint $V_N$ is promoted to a 1-parameter family of unitary
operators. Varying $N^a$, we obtain, on $\Ha$, a unitary
representation of the group of diffeomorphisms on $\Sigma$ generated
by complete analytic vector fields. Thus, there are no anomalies.
Note that this is {\it not} a formal argument; the operators $U(t)$
corresponding to $V_N$ are rigorously defined on a proper Hilbert
space, and they are unitary because the measure $\mu_0$ is
diffeomorphism invariant.

Note that $U_\varphi$ preserves the space $\Cyl^\infty(\agb)$ of
smooth cylindrical functions.  Since $\Cyl^{\infty}(\agb)$ is also
preserved by our algebra of elementary quantum operators (generated by
$\hat{T}_\alpha$ and $\hat{T}_S$), it is natural to take
$\Cyl^\infty(\agb)$ to be the dense subspace $\Phi \subset \haux$ of
step 5$'$b of the refined algebraic quantization program.  Finally, we
need to specify a topology on $\Phi$. Finite dimensional examples
suggest that we let one of the standard nuclear topologies of the
$C^\infty(\A_\g) \cong C^\infty(SU^n(2))$ induce the required topology
on $\Cyl^{\infty}(\agb)$.

We will seek `solutions of the constraints' in the topological dual
$\Ps$, the space of cylindrical distributions.  Diffeomorphisms have a
natural action on $\o \phi \in \Ps$ by duality and we will say that
$\o \phi \in \Phi'$ is a solution of the diffeomorphism constraints if
\be
\o \phi(U_\varphi\circ \phi) = \o\phi (\phi)
\qquad {\rm for \ all} \ \ \varphi
\in {\rm Diff }(\S) \ {\rm and} \ \phi \in \Phi \ .
\ee
Many such distributions exist. For example, given any spin-network
state $|\alpha, \vec \pi, \vec c\r$ we may define a distribution
$\mu_{\alpha,\vec \pi,\vec c}$ through its action on any $\phi
\in \Cyl^\infty(\agb)$:
\be
\label{DinvM}
\mu_{\alpha, \vec \pi,\vec c} [\phi]\ := \
\sum_{|\alpha_2, \vec \pi_2,\vec{c_2}\r  \in
[|\alpha, \vec \pi,\vec c\r]}
\l \alpha_2 \vec \pi_2,\vec c_2 | \phi \r
\ee
where $[|\alpha, \vec \pi,\vec c\r]$ is the set of all spin-network
states $|\alpha_2, \vec \pi_2,\vec c_2 \r$ such that $U_\varphi
|\alpha, \vec \pi,\vec c \r = |\alpha_2, \vec \pi_2,\vec c_2\r$ for
some $\varphi \in {\rm Diff} (\Sigma)$. To see that this sum converges
and $\mu_{\alpha, \vec \pi ,\vec c}$ is a well defined element of
$\Phi'$, write $\phi$ as
\be
\label{star}
\phi \ =\ \sum_{\g',\vec{\pi} } f_{\g',\vec{\pi}' }
\end{equation}
where $ f_{\g',\vec{\pi}}$ is the orthogonal projection of $\phi$ onto
the space $\Hva$. The sum ranges over all the vector spaces of the
orthogonal decomposition, however, since $\phi$ is cylindrical, there
are contributions only for $\g'\le\g$ for some graph $\g$.  Substitute
(\ref{star} ) into (\ref{DinvM}). On the right hand side, the products
vanish unless $(\a_2,\vec{\pi_2})=(\g', \vec{\pi}')$.  Thus, there are
only a finite number of nonzero terms; the right hand side of
(\ref{star}) is finite and defines an element of $\Phi'$.  Note that,
heuristically, we have invoked the idea of group averaging to
construct these distributions, using a discrete measure on the orbit
of $|\alpha, \pi, c\r$ under ${\rm Diff}(\Sigma)$.

\end{subsection}

\begin{subsection} {The issue of independent sectors}
\label{srarg}

Having identified a suitable dense subspace $\Phi \subset \haux$ and
having seen that its topological dual $\Ps$ is large enough to contain
diffeomorphism invariant distributions, we now wish to construct a map
$\eta: \Phi \rightarrow \Ps$ that completes step 5$'$c in our program.
This will, however, be more complicated than for the examples in
Sec.II due to the fact that each state $|\phi\r \in \Phi$ has an
infinite `isotropy group' of diffeomorphisms that leave $|\phi\r$
invariant.  Thus, the sum in (\ref{DinvM}) was not over the entire
diffeomorphism group, but only over the {\it orbit} of the state
$|\alpha, \pi, c \r \in \Phi$.

Because the sum in (\ref{DinvM}) itself depends on the state $|\alpha,
\vec \pi, \vec c\r$, our definition of the inner product on $\vd$ will
have to take into account the fact that the orbit size is
state-dependent.  While the infinite size of the orbits would appear
to make this difficult, a simplification will occur as the presence of
`infinitely different' isotropy groups will imply that
$L^2(\agb,d\mu_0)$ carries a {\it reducible} representation of the
algebra of observables.  In fact, we show below that $\haux$ can be
written as a direct sum of subspaces such that, on each subspace, the
sizes of orbits are `comparable'.  This will allow us to give a well
defined averaging procedure by treating each such subspace separately
in section \ref{strat}.  A similar situation is discussed in appendix
\ref{supsel}.

In order to classify these isotropy groups, let us consider for each
spin-network state $|\alpha, \vec \pi, \vec c\r$ the collection ${\cal
E}_\alpha$ of analytic edges of the graph $\alpha$.  For technical
reasons, we shall focus on graphs for which, given any edge $e \in
{\cal E}_\alpha$, there is an analytic real function $f$ which
vanishes on the maximal analytic curve $\tilde e$ that extends $e$,
but nowhere else.  We shall call such graphs (and their associated
curves) `type I', while all others are `type II.'  Note that the
collection of $\tilde e$ defined by the type I graph $\alpha$
intersect at most a countable number of times and so define a graph
$\tilde\alpha$ with countably many edges.

Now, given any $n$ type I maximal analytic curves (i.e., curves which
cannot be analytically extended) in $\Rl^3$ and any distinct maximal
analytic curve $\tilde e$ (not necessarily of type I), there is a
multi-parameter family of analytic diffeomorphisms that preserves the
$n$ type I curves but does not preserve $\tilde e$.  To see this,
begin with any constant vector field $X_0$ on $\Rl^3$ which is not
everywhere tangent to $\tilde e$ on $\tilde e$.  Let $f_i$ be the real
analytic function that vanishes exactly on the $ith$ maximal type I
curve.  Then the product $f$ of the the $f_i$ is a real analytic
function that vanishes exactly on the union of these curves.  Thus,
the complete analytic vector field $X = f e^{-f^2} X_0$ exponentiates
to a one parameter family of analytic diffeomorphisms that preserves
the $n$ maximal type I curves, but does not preserve $\tilde e$.

Thus, for two spin-network states $|\alpha_1, \vec \pi_1,\vec c_1\r$
and $|\alpha_2, \vec \pi_2, \vec c_2 \r$, (with $\alpha_1$ of type I)
either $\tilde \alpha_1$ and $\tilde \alpha_2$ are identical or there
are infinitely many diffeomorphisms $\varphi$ which preserve one of
these (say, $\tilde \alpha_1$) but move the other (say, $\tilde
\alpha_2$).

As in section \ref{s3}, we consider the algebra $\bsp$ of operators
$A$ on $\haux$ that $i)$ are defined on $\Phi$ and map $\Phi$ into
itself, $ii)$ have adjoints $A^\dagger$ defined on $\Phi$ which map
$\Phi$ into itself, and $iii)$ commute with the action of all
diffeomorphisms $\varphi$. (Note that the last condition implies that
$A^\dagger$ also commute with constraints.)  Let $|\phi_1\r =
|\alpha_1,\vec \pi_1,\vec c_1\r$ and $|\phi_2 \r = |\alpha_2,\vec
\pi_2,\vec c_2\r$ be the spin-network states above, so that there are
infinitely many diffeomorphisms $\varphi$ which move $\tilde \alpha_2$
but move no edge of $\tilde \alpha_1$.  Then, for such a $\varphi$,
the matrix elements $\l \phi_1|A|\phi_2\ra$ of any $A \in \bsp$ must
satisfy
\be \l \phi_1
|A|\phi_2 \ra = \l \phi_1|\varphi A | \phi_2 \ra = \l \phi_1|A \varphi
|\phi_2 \ra
\ee
while $\l \phi_2| \varphi | \phi_2 \ra = 0$.  Thus, either $\l
\phi_1|A|\phi_2 \r_{aux} = 0$ or the vector $A^\dagger|\phi_1\r$ has
an infinite number of equal components.  However, $|\phi_1\r \in \Phi$
lies in the domain of $A^\dagger$ so that $A^\dagger|\phi_1\r$ is
normalizable, whence $\l \phi_1 |A|\phi_2 \ra $ must vanish.  Since
the adjoint of $A$ is also in $\bsp$, $\l \phi_2 |A| \phi_1 \ra^* $
vanishes as well.  We thus have a `super-selection rule' between
states associated with the graphs $\tilde \alpha_1 $ and $\tilde
\alpha_2$ and the representation of $\bsp$ on $\haux$ is reducible.

Note in particular that we have a superselection rule between states
associated with graphs of type I and type II.  We may thus decompose
the representation of $\bsp$ on $L^2(\agb,d\mu_0)$ as $\haux = \haux^I
\bigoplus \haux^{II}$.  Because the subspace $\haux^I$ is technically
much simpler, we will focus on this sector and ignore $\haux^{II}$ in
what follows. (This division into two types would be unnecessary if we
could replace the analytic loops by smooth ones in the begining of
our construction. See Sec. \ref{diss}.)

Our discussion above implies that $\haux^I$ is in fact a direct sum of
representations, each acting in a subspace $\haux^{\tilde \alpha}$
associated with a given (maximally analytically extended) graph
$\tilde \alpha$.  On the other hand, a diffeomorphism $\varphi$ will
map one such subspace $\haux^{\tilde \alpha}$ to another
($\haux^{\tilde {\varphi \alpha}}$).  It is therefore convenient to
consider the class $[\tilde \alpha]$ of all maximally extended
analytic graphs $\tilde \alpha_1$ which can be mapped onto $\tilde
\alpha$ by an analytic diffeomorphism.  This gives a decomposition of
$\haux$ through \be \haux^I = \bigoplus_{[\tilde \alpha]}
\haux^{[\tilde \alpha]}, \ee where \be \haux^{[\tilde \alpha]} =
\bigoplus_{\tilde \alpha_1 \in [\tilde \alpha]} \haux^{\tilde
\alpha_1} \ee where both direct sums are implicitly over only graphs
of type I.  The sectors $\haux^{[\tilde \alpha ] }$ are truly
independent in the sense that they are not mixed by any physical
operators $A \in \bsp$ or any diffeomorphism $\varphi$.  Thus, from
now on, we will treat each $\haux^{[\tilde \alpha]}$ individually.

\end{subsection}

\begin{subsection}{A Family of Maps}
\label{strat}

We now wish to implement step 5$'$ of the program separately within
each `independent sector' $\haux^{[\tilde \alpha]}$ of
$L^2(\agb,d\mu_0)$, associated with the class of maximally
analytically extended graphs $\tilde \alpha_1$ which are mapped onto
$\tilde \alpha$ by analytic diffeomorphisms.  Thus, for each
$\haux^{[\tilde \alpha]}$, we introduce the dense subspace
$\Phi^{[\tilde \alpha]} \subset \haux^{[\tilde \alpha]}$ of functions
that are ($C^\infty$) cylindrical over graphs associated with $[\tilde
\alpha]$ and the corresponding topological dual $\Pgs$.  We will
identify a vector space $\vd^{[\tilde \alpha]}$ and impose an inner
product to define a Hilbert space $\hd^{[\tilde \alpha]}$ of
diffeomorphism invariant states.  As before, $\vd^{[\tilde \alpha]}$
will be the image of a family of maps $\eta^{[\tilde \alpha]}_{a}:
\Phi^{[\tilde \alpha]} \rightarrow \Pgs$ and will contain only
diffeomorphism invariant distributions. (Here, $a = a_{[\tilde
\alpha]} \in \Rl^+$. For simplicity of notation, we will not make the
dependence of $a$ on $[\tilde\alpha]$ explicit.)

To construct the map $\eta^{[\tilde \alpha]}_a$, let us first give its
action on functions $|f\r \in \bigoplus_{\vec \pi} \Hva$ associated
with some fixed graph $\gamma$ with $\tilde \gamma \in [\tilde
\alpha]$.  The action of $\eta^{[\tilde \alpha]}_{a}$ on general
states $|\phi\r \in \Phi$ then follows by (finite) anti-linearity.  To
construct this map, we will need to consider the `isotropy' group
$Iso(\tilde \g)$ of diffeomorphisms that map $\tilde \g$ to itself and
the `trivial action' subgroup $TA(\tilde \g)$ of $Iso(\tilde \g)$
which preserves each edge of $\tilde \g$ separately.  We will also
need the quotient group $GS(\tilde \g) = Iso(\tilde \g)/TA(\tilde \g)$
of `graph symmetries' of $\tilde \g$ and some set $S(\tilde \g)$ of
analytic diffeomorphisms which has the property that, for all $\tilde
\beta \in [{\tilde \g}]$ there is exactly one $\varphi \in S(\tilde
\g)$ that maps $\tilde \g$ to $\tilde \beta$. The appropriate maps
are then given by
\be
\label{loopeta}
\eta_a^{[\tilde{\alpha}]} |f\r =
a\  \bigl( \sum_{\varphi_1 \in S(\tilde \g)}
\sum_{[\varphi_2] \in GS(\tilde \g)} \varphi_1 \varphi_2 |f \r
\bigr)^\dagger
\ee
where, in the second sum, $\varphi_2$ is any diffeomorphism in the
equivalence class $[\varphi_2]$.  For the reader who feels that this
definition has been `pulled out of a hat,' we will provide a heuristic
`derivation' below in section \ref{heur} by `renormalizing' the map
given by naive group averaging.

In order to show that $\eta_a^{[\tilde{\alpha}]} |f \r$ does in fact
define an element of $\Phi'$, note that its action on any state $|g \r
\in \bigoplus_{\vec \pi} {\cal H}_{aux}^{\beta, \vec \pi}$ is given by
\be
\label{finite}
\bigl( \eta_a^{[\tilde{\alpha}]} |f \r \bigr)
\bigl[ |g \r \bigr]
=  \delta_{[{\tilde \g}],[{\tilde \beta}]}
\sum_{\varphi_2 \in GS(\tilde \g)}
\l f | \varphi_2 \varphi_0 | g \r
\ee
where $\varphi_0$ is any diffeomorphism that maps $\tilde \g$ to
$\tilde \beta$.  Because $\tilde \g$ may have an infinite number of
edges, $GS(\tilde \g)$ may be infinite as well.  Nonetheless, we will
now show that the above sum contains only a finite number of nonzero
terms.

First, note that if there are any nonzero terms at all, we may take
$\varphi_0 \g = \beta$ without loss of generality.  In this case, a
term in (\ref{finite}) is nonzero only if the associated $\varphi_2$
preserves that graph $\g$.  The key point is to note that, since
$\tilde \g$ may be constructed by analytically extending the edges of
the graph $\g$, the action of any analytic diffeomorphism on the edges
of $\g$ determines the action of this diffeomorphism on every edge in
the {\it extended} graph $\tilde \g$.  Thus, the diffeomorphisms
$\varphi_2 \in S(\tilde \g)$ that preserve $\tilde \g$ must rearrange
the edges of $\g$ in distinct ways.  Since $\g$ contains only a finite
number of edges, it follows that there can be at most a finite set of
diffeomorphisms $\varphi_2$ in $GS(\tilde \g)$ that preserve $\g$.
There are thus only finitely many terms in (\ref{finite}).  The fact
that (\ref{loopeta}) defines an element of $\Phi'$ then follows by
(finite) linearity.

The space $\vd^{[\tilde \alpha]}$ is then defined to be the image of
$\eta^{[\tilde \alpha]}_a$.  It is clear from the form of the sum
(\ref{finite}) that $\eta^{[\tilde \alpha]}_a$ is real and positive so
that the inner product (\ref{finite}) is well-defined, Hermitian, and
positive definite.  We may therefore complete each $\vd^{[\tilde
\alpha]}$ to define a Hilbert space $\hd^{[\tilde \alpha]}$.

Furthermore, $\eta_a^{[\tilde \alpha]}$ commutes with all $A$ in
$\bsp$ in the sense that
\ba
(\eta_a^{[\tilde \alpha]} \phi_1)[A \phi_2] &=& \sum_{\varphi
\in GS(\tilde \alpha)} \l\phi_1|\varphi A|\phi_2\r \nonumber\\
&= &\bigl( \eta_a^{[\tilde \alpha]} (A^\dagger \
\phi_1)  \bigr)[\phi_2].
\ea
(Here, without loss of generality, we take $\phi_1, \phi_2$
cylindrical over $\tilde \alpha$.)  It follows that the map $A \mapsto
A_{phys}$ (where $A_{phys}(\eta \phi) = \eta( A \phi)$ ) defines an
anti $\star$-representation of $\bsp$ on $\hd^{[\tilde
\alpha]}$. Thus, the ``reality conditions'' on physical observables
have been incorporated.

\end{subsection}
\begin{subsection} {Some final Heuristics}
\label{heur}

For those who are interested, we now present a short heuristic
`derivation' of (\ref{finite}) in which we first average over the
entire group of diffeomorphisms (in analogy with \cite{AH2,AH3} and
section \ref{ex}) and then `renormalize' the resulting distribution by
canceling (infinite) volumes of isotropy groups.  Because a sum of the
form $\sum_{\varphi \in {\rm Diff}(\Sigma)} \varphi |\phi\r$ diverges
(even as an element of $\Ps$), we attempt to remove this divergence by
comparing the inner product of two distributions $\o \psi$ and $\o
\psi$ in $\vd^{[\tilde \alpha]}$ with the norm of some reference
distribution $\o \rho$ which lies in the same vector space
$\vd^{[\tilde \alpha]}$.  Let us suppose that these `heuristic
distributions' are obtained by averaging $|\phi\r, |\psi\r$, and
$|\rho\r \in \Phi^{[\tilde \alpha]}$ over the diffeomorphism group.
For convenience, we will also fix some particular extended analytic
graph $\tilde \alpha$ and assume that $|\phi\r$, $|\psi\r$, and
$|\rho\r$ lie in $\haux^{\tilde \alpha}$.  Then, the {\it ratio} of
the inner product of $\o \phi$ and $\o \psi$ to the norm of $\o \rho$
is
\be
\label{hq}
{{\l \o \phi | \o \psi\rd} \over {\l \o \rho|\o \rho\rd}}
= {{\sum_{\varphi \in {\rm Diff}(\Sigma)} \l
\psi | \varphi |\phi \r}\over
{\sum_{\varphi \in {\rm Diff}(\Sigma)} \l \rho| \varphi | \rho \r}}
\ee
so that a given diffeomorphism $\varphi$ contributes to this sum only
if it preserves $\tilde \alpha$.  That is, we need only sum over the
isotropy group $Iso(\tilde \alpha)$.  Note that we may rewrite the
sums over diffeomorphisms in (\ref{hq}) as sums over the cosets
$GS(\tilde \alpha )$ (which give a finite result by the above
discussion) multiplied by the (infinite) size of the trivial action
subgroup $TA(\tilde \alpha)$.  Formally canceling these infinite
factors in the numerator and denominator, we arrive at

\be {{\l \o \phi|\o \psi\rd} \over {\l\o \rho|\o\rho\rd}} =
{{\sum_{[\varphi] \in GS(\tilde \alpha)} \l \psi| \varphi|\phi
\r}\over {\sum_{[\varphi] \in GS(\tilde \alpha)} \l \rho| \varphi |
\rho \r}},
\ee
where the sum is over the equivalence classes in $GS(\tilde \alpha)$
and $\varphi$ is an arbitrary representative of $[\varphi]$.  This
motivates the definition (\ref{loopeta}) of the maps $\eta^{[\tilde
\alpha]}_a$ and the inner product (\ref{finite}).

\end{subsection}

\begin{subsection} {Subtleties}
\label{subtle}

We have seen that the Hilbert space $\hd$ that results from solving
the (Gauss and the) diffeomorphism constraints can be decomposed as a
direct sum of Hilbert spaces $\hd^{[\tilde\alpha]}$, each of which
carries a representation of the algebra $\bsp$ of physical operators.
Thus, elements of $\bsp$ --observables which strongly commute with
constraints-- do not mix states from distinct Hilbert spaces that
feature in the direct sum. Recall, however, that the physical
observables have to commute with constraints only weakly and there may
well exist weak observables which connect distinct Hilbert spaces.
 From a physical viewpoint, therefore, we need to focus on the
irreducible representations of the algebra of weak observables.  If
there are no further constraints, (as in the Husain-Kucha\v{r} model),
these irreducible sectors are properly thought of as separate and, in
the standard jargon, {\it super-selected}. Since by assumption the
Hamiltonian operator commutes with all constraints, dynamics will
leave each sector invariant.  Indeed, no physical observable can map
one out of a superselected sector.  Thus, a physical realization of
the system will involve only one such sector and just which sector
arises must be determined by experiment. Unfortunately, as the matter
stands, we do not have a manageable characterization of these sectors
because we focussed only on strong observables. (In general, weak
observables do not satisfy (\ref{stareq}).)

If the diffeomorphism group represents only a sub-group of the full
gauge group (as in the case of general relativity), then there can be
a further complication and the situation becomes quite subtle. On the
one hand, because we have more constraints, one expects there to be
fewer observables. On the other hand, commutator of the an
operator with the diffeomorphism constraints may be equal to one of
the new constraints. Then, while the operator would not be an
observable of the partial theory that ignores the additional
constraints, it {\it would} be an observable of the full theory.
Curiously, this is precisely what happens in the case of
3-dimensional, Riemannian general relativity (i.e., the $ISU(2)$
Chern-Simons theory). The Wilson loop operators $\hat{T}_\alpha$ fail
to be weak observables if we consider only the diffeomorphism
constraint but they {\it are} weak observables of the full
theory. Furthermore, they mix the independent sectors which are
super-selected with respect to diffeomorphisms. We expect that the
situation will be similar in 4-dimensional general relativity. Thus,
we expect that the physical states of this theory will not be confined
to lie in just one $\hd^{[\wh\alpha]}$; as far as general relativity
is concerned, one should not think of these sectors as being {\it
physically} super-selected.

Finally, note that we have asked that the physical states be invariant
only under diffeomorphisms generated by vector fields. Large
diffeomorphisms are unitarily implemented in the physical Hilbert
space; they are symmetries of the theory but not gauge. One {\it may}
wish to treat them as gauge and ask that the ``true'' physical states
be invariant under them as well. If so, one can again apply the group
averaging procedure, now treating the modular group as the gauge
group.  In the case of 3-dimensional Riemannian general relativity on
a torus, for example, this procedure is successful and yields a
Hilbert space of states that are invariant under all diffeomorphisms.

\end{subsection}

\end{section}

\begin{section} {Discussion}
\label{diss}

In this paper, we have presented solutions to the Gauss and the
diffeomorphism constraints for a large class of theories of
connections. The reader may be concerned that we did not apply the
quantization program of Sec. II to the Gauss constraint but instead
solved it classically. However, we chose this avenue only for brevity;
it is straightforward to first use the program to solve the Gauss
constraint and then face the diffeomorphism constraint. In this
alternate approach, one begins with the space $\ab$ of generalized
connections (see section 4.2) as the classical configuration space and
lets the auxiliary Hilbert space be $L^2(\ab, d\mu'_0)$, where
$\mu'_0$ is the induced Haar measure on $\ab$ (see Ref. [7,12]).
Next, one introduces the Gauss constraints as operators on the new
auxiliary Hilbert space. The resulting unitary operators just
implement the action of the group $\Gb$ of generalized gauge
transformations on the Hilbert space.  Since $\Gb$ is compact, the
resulting group averaging procedure is straightforward and leads us to
$L^2(\agb, d\mu_0)$ as the space of physical states with respect to
the Gauss constraints.  One is now ready to use Sec. \ref{phy} to
implement the diffeomorphism constraints.

The final picture that emerges from our results can be summarized as
follows.  To begin with, we have the auxiliary Hilbert space $\Ha$.
While it does not appear in the final solution, it does serve three
important purposes. First, it ensures that real, elementary functions
on the classical phase space are represented by self-adjoint
operators, so that the ``kinematical reality conditions'' on the full
phase space are incorporated in the quantum theory. Second, it enables
us to promote constraints to {\it well-defined} operators thereby
making the analysis of potential anomalies mathematically
sound. Finally the space $\Phi$, whose topological dual $\Phi'$ is the
``home'' of physical quantum states, is extracted as a dense sub-space
of $\Ha$. The physical states $\bar\phi\in \Phi'$ are obtained by
``averaging'' states $\phi\in \Phi$ over the orbits of the
diffeomorphism group appropriately. Care is needed because the orbits
themselves have an infinite volume and because, in general, different
orbits have different isotropy groups. These features lead to
diff-superselected sectors. Each sector is labeled by the
diffeomorphism class $[\tilde \alpha]$ of ``maximally extended'' (type
I) graphs $\tilde \alpha$.  Operators on $\Ha$ which leave $\Phi$
invariant have an induced action on the topological dual, $\Phi'$ of
$\Phi$. If they commute with the diffeomorphism operators on $\Ha$,
they descend to the space $\vd$ of (diff-)physical states. The sectors
are diff-superselected in the sense that each of them is left
invariant by operators on $\Phi'$ which descend from observables
--i.e., self-adjoint operators which commute with the
diffeomorphism operators-- on $\Ha$.  The induced scalar product on
${\cal V}_{\rm diff}$ is unique up to an overall multiplicative
constant on each diff-superselected sector.  It automatically
incorporates the physical reality conditions. (The ambiguity of
multiplicative constants would be reduced if there exist weak
observables which mix these sectors which are superselected by strong
observables.)

How does this situation compare to the one in the general algebraic
quantization program of Ref. [17,18]? In the final picture, the inner
product {\it is} determined by the reality conditions.  However, the
group averaging strategy enables one to find this inner product {\it
without} having to find the physical observables explicitly; the inner
product on $\Ha$ which incorporates the kinematical reality conditions
on the full phase space descends to $vp$.  This is an enormous
technical simplification. On the conceptual side, there are now four
inputs into the program: choice of a set of elementary functions
(labeled by loops and strips in our case), of a representation of the
corresponding algebra (on $L^2(\agb, \mu_0)$ in our case), of
expressions of the regularized constraint operators (which, in our
case, implement the natural action of the diffeomorphism group on
$\Ha$), and of the subspace $\Phi$ ($\Cyl^\infty(\agb)$ in our case).
We have shown that the choices we made are viable and quantization can
be completed.  There may of course be other, inequivalent quantum
theories, which correspond to different choices.  Indeed, even in
Minkowski space, a classical {\it field} theory can be quantized in
inequivalent ways. We expect, however, that there exists an
appropriate uniqueness theorem which singles out our solution,
analogous to the theorem that singles out the Fock representation for
free field theories.

What are the implications of these results to the specific models
discussed in Sec. III?  For the Husain-Kucha\v{r} model, we have
complete solutions. For Riemannian general relativity, on the other
hand, we have only a partial result since the Hamiltonian constraint
is yet to be incorporated. However, our analysis does provide a
natural strategy to complete the quantization.  For, we already have
indications that the projective methods can be used also to regulate
the Hamiltonian constraint operator {\it on diffeomorphism invariant
states}. If this step can be completed, one would check for
anomalies. If there are none, one would again apply the group
averaging procedure to find solutions. This task may even be simpler
now because, given the structure of the classical constraint algebra,
one would expect the Hamiltonian constraints to {\it commute} on
diffeomorphism invariant states. The procedure outlined in Appendix A
would then lead to the physical Hilbert space for the full theory. As
indicated in Sec. \ref{subtle}, however, subtleties will arise because of the
observables which commute with the constraints only weakly and the
final Hilbert space is likely to contain elements from {\it different}
diff-superselected sectors. Furthermore, to extract ``physical''
predictions, one would almost certainly have to develop suitable
approximation schemes. However, this task would be simplified
considerably if we already know that a consistent quantum theory
exists. Indeed, in this respect, the situation would be comparable to
the one currently encountered in atomic and molecular physics where
approximations schemes are essential in practice but the knowledge
that the exact Hamiltonian exists as a well-defined self-adjoint
operator goes a long way in providing both confidence in and
guidelines for these approximations.

For Lorentzian general relativity, one can begin with the formulation
in which the spin connection is the configuration variable.  For this
case, the results of this paper again lead to a complete solution to
the Gauss and the diffeomorphism constraints. Unfortunately, as
mentioned in the Introduction, the Hamiltonian constraint is
unmanageable in these variables and the best strategy is to perform a
transformation and work with {\it self-dual} connections
\cite{2}. Classically, the required canonical transformation is
well-understood. Its quantum analog is an appropriate ``coherent state
transform'' which would map complex-valued functions of spin
connections to holomorphic functions of the self-dual
connections. Such a transform is already available
\cite{6} and it seems fairly straightforward to carry over our
treatment of the diffeomorphism constraint to the holomorphic
representation. However, it is far from being obvious that the
Hamiltonian constraint can be treated so easily in the holomorphic
representation. Another strategy is to begin with the Riemannian
model, obtain physical states and {\it then} pass to the holomorphic
representation via an appropriate generalization of the Wick rotation
procedure. Thus, whereas in the Riemannian case, results of this paper
provide a clear avenue, in the Lorentzian case, new inputs
are needed. Work is in progress along the two lines indicated above.

The canonical approach to quantum gravity is quite old; foundations of
the geometrodynamic framework were laid by Dirac and Bergmann already
in the late fifties. The precise mathematical structure of the
classical configuration and phase spaces became clear in the
seventies.  However, these analyses dealt only with smooth fields
while, as is well-known, in quantum field theory one has to go beyond
such configurations.  The required extensions are non-trivial and are,
in fact, yet to be carried out in the metric representation.
Consequently, in the traditional geometrodynamical approach, the
formulation and imposition of quantum constraints have remained at a
formal level {\it even for the diffeomorphism constraint}. We have
seen that the situation changes dramatically if one shifts the
emphasis and works with connections.  (Note that these can be $SU(2)$
spin connections; they don't have be self-dual. Since the spin
connection is completely determined by the triads, the corresponding
representation provides an alternative framework to solve the quantum
Gauss and diffeomorphism constraints of the triad geometrodynamics.)
Now, problems of quantum field theory can be faced directly and the
general level of mathematical precision is comparable to that
encountered in rigorous quantum field theory. Finally, note that this
became possible {\it only} because of the availability of a calculus
on the quantum configuration space which does not refer to a
background field such as a metric. Thus, the projective techniques
summarized in Sec. \ref{s4} are not a luxury; they are essential if
one wants to ensure that inner products and operators are well-defined
in the quantum theory.

Most of theoretical physics, however, does not require such a high
degree of precision. Why, then, is so much care necessary here?  The
main reason is that we have very little experience with
non-perturbative techniques. We have already seen that the
perturbative strategy, which is so successful in theories of other
forces of Nature, fails in the case of gravity. Hence, if one wishes
to pursue a new approach, it is important to have an assurance that
the quantum theory we are dealing with is internally consistent
and that the problems that arise in perturbative treatments are not
just swept under a rug. An obvious way to achieve certainty is to
work at a high level of mathematical precision.

The mathematical framework could, however, be improved in two
directions. First, the functional calculus we used is based, in an
essential way, on the assumption that all edges of our graphs are
analytic. If we weaken this assumption and allow edges which are only
$C^\infty$, a number of technical problems can arise since, for
example, two $C^\infty$ curves can have an infinite number of
intersections in a finite interval. On physical grounds, on the other
hand, smoothness seems more appropriate than analyticity and it would
be desirable to extend this framework accordingly. Furthermore, if we
could work with smooth loops, the discussion of the ``independent
sectors'' in Sec. \ref{phy} would simplify considerably; it would not
be necessary to divide the spin networks into types.  The second
improvement would be more substantial. The present mathematical
framework is based on the assumption that traces of holonomies should
become well-defined operators on the auxiliary Hilbert space. Once
this assumption is made, one is naturally led to regard $\agb$ as the
quantum configuration space and use on it the calculus that is induced
by the projective techniques. The assumption is not unreasonable for a
diffeomorphism invariant theory and has led to a rich structure which,
as we saw, is directly useful in a number of models. (The framework
has also been used to find new results in 2-dimensional Yang-Mills
theory \cite{ALMMT1} which happens to be invariant under all volume
preserving diffeomorphisms.)  However, it is quite possible that,
ultimately, the assumption will have to be weakened. To do so, we may
need to feed more information about the underlying manifold into the
quantum configuration space. Our present construction does capture a
part of the manifold structure through its use of analytic graphs and
also has some topological information, e.g., of the first homotopy
group of the manifold. However, it does not use the notion of
convergence of a sequence of graphs which knows much more about the
topology of the underlying manifold.  In the language of projective
techniques (see Appendix \ref{proj}), it would be desirable to use the
underlying manifold to introduce a topology on the label set and see
how it influences the rest of the construction. These issues are
currently being investigated.

\end{section}

{\centerline {\bf Acknowledgments}}

We would like thank Rudolfo Gambini, Peter Peldan, Jorge Pullin, Carlo
Rovelli and Lee Smolin for discussions and are especially grateful to
John Baez for his valuable inputs.  AA and TT were supported in part
by the NSF Grant PHY93-96246; JL was supported in part by the KBN
grant 2-P302 11207; DM was supported in part by the NSF grant
PHY90-08502; and JM was supported in part by JNICT grant
STRDA/C/PRO/1032/93. In addition, the Eberly fund of the Pennsylvania
State University provided partial support to enable DM and JM to visit
the Center for gravitational physics and geometry where this work was
completed.

\appendix

\begin{section} {Super-selection Rules for Abelian Constraints}

\label{supsel}

To illustrate the quantization program, we discussed a number of
simple  examples in section \ref{ex}. To bring out some subtleties
associated with the group averaging procedure, in this Appendix, we
will consider a somewhat more general situation which, however, is
simpler than the one considered in Sec. \ref{phy}.

In section \ref{ex}, the group generated by the quantum constraints
was Abelian and was represented by unitary operators $U(g)$ in a
Hilbert space $\haux$.  The definition of the physical inner product
involved a map $\eta$ from a space $\Phi$ of test functions to its
topological dual $\Ps$ which was defined by integrating over the
volume of the gauge group, $\eta |f\r = (\int dg U(g) |f\r)^\dagger$.
As such, it is clearly important that no infinite subgroup should
leave $|f\r$ invariant so that the integral does not diverge.  Thus,
it is natural to ask if this method can be suitably modified to
incorporate the case when some $U(g)$ have eigenstates in $\haux$ with
eigenvalue 1. In this Appendix, we will analyze this issue in the
general setting of Abelian constraints and show that the answer is
always `yes,' though the procedure is somewhat more subtle.

Recall that our intent is to construct an irreducible representation
of a $\star$-algebra of {\it physical} operators and that we suppose
this algebra to be represented on $\haux$.  At least when this algebra
is generated by bounded operators, we will see that the representation
on $\haux$ is reducible whenever $1$ is a part of the discrete as well
as the continuous spectrum of some $U(g)$.

Suppose that the representation of the gauge group is generated by
some set $U_i$ of unitary operators for $i$ in some label set
$I$. Denote by $S_i^{(d)}$ the subspace of $\haux$ which is left
invariant by $U_i$, i.e., the space of eigenvectors with discrete
eigenvalue $1$. Since $\{1\}$ is a set of zero measure in $\Rl$, any
state in $\haux$ which is orthogonal to ${\cal S}^1_i$ can be built
from spectral subspaces of $U_i$ with eigenvalue $\not= 1$.  Now,
solutions to the constraints in $\Phi'$ are of two types. First, each
element of ${\cal S}^d_i$, regarded as an element of $\Phi'$, is a
solution. Second, there is a subspace ${\cal S}^c_i$ obtained by
group-averaging elements of $\Phi$ which are orthogonal to ${\cal
S}^d_i$.  These two subspaces of physical states are orthogonal to
each other. Consider now a bounded operator $A$ which commutes with
each $U_i$.  It is straightforward to check that the action of $A$
preserves each of the two orthogonal subspaces; the action of $A$ on
$\Phi'$ does not mix the discrete and continuous eigenvalue $1$
distributions of $U_i$ in $\Ps$.

We now refine our group-averaging procedure as follows.  First,
decompose $\haux$ as a direct sum of the subspaces $\haux^\lambda$,
where $\lambda$ is a map $\lambda: I \rightarrow \{d,c\}$.  Thus,
$\haux^\lambda$ is the subspace on which $U_i$ has continuous spectrum
for $\lambda(i) = c$ but has discrete spectrum for $\lambda(i) = d$.
Since these subspaces are superselected, it is only meaningful to
define a physical Hilbert space $\hp^\lambda$ for each $\haux^\lambda$
separately.  This is done by projecting $\haux^\lambda$ to the zero
spectrum of each $U_i$ with $\lambda(i) = d$ and averaging as in
section \ref{ex} over the Abelian group generated by the $U_i$ with
$\lambda(i) = c$.  It then follows that operators induced by physical
operators on $\haux$ have the required $\star$-relations on each
$\hp^\lambda$.

We would like to emphasize that, when the $U_i$'s generate the entire
gauge group, these super-selection rules are not just an artifact of
the mathematics but are important for a physical understanding of the
system.  They imply that the representation of the physical algebra on
$\haux$ is reducible, so that each $\hp^\lambda$ contains a separate
representation of the algebra of physical operators.  Which
$\hp^\lambda$ is realized in a given situation must be determined
experimentally.

Furthermore, the super-selection rules described above have a close
classical analogue due to Liouville's theorem.  Consider a classical
constraint function $C_i$ and a strong observable $A$ that is a smooth
function on the phase space. (The use of strong observables is not
essential but simplifies the discussion.)  The Hamiltonian vector
field $h_A$ of any such $A$ has the property that it maps any orbit of
$C_i$ in the unconstrained phase space onto another orbit of $C_i$.
Heuristically, regions of the phase space that contain compact orbits
correspond to the discrete spectrum of a corresponding $U_i$ and
regions that contain non-compact orbits correspond to the continuous
spectrum.  Now, consider any set of compact orbits with non-zero but
finite phase space volume.  By Liouville's theorem, the exponentiated
action of any Hamiltonian vector field preserves the finite volume of
this set.  As a result, $h_A$ cannot map this set of compact orbits to
a bundle of non-compact orbits.  Note that this is a direct analogy
with the super-selection laws described above.

Of course, if these orbits are not the full gauge orbits, but only
those of a gauge subgroup, then such arguments are inconclusive
when applied to weakly physical operators.
This is because, under
the action of the full gauge group, many of the above compact
orbits may combine to form a single non-compact orbit, which could
then be mapped onto non-compact orbits in a volume preserving way.

\end{section}

\begin{section} {Projective Limits}
\label{proj}

A general setting for functional integration over an infinite
dimensional, locally convex, topological space $V$ is provided by the
notion of ``projective families'' \cite{Ya,DWM}.  This framework can
be naturally extended to theories of connections where the relevant
space $\agb$ is {\it non-linear}\cite{AL1,L,MM,AMM,AL2,AL3}. In the
present appendix we will summarize the basic ideas which are
implicitly used in the main text.

Let $L$ be a partially ordered directed set; i.e. a set equipped
with a relation `$\ge$' such that, for all $S, S'$ and $S''$ in $L$
we have:

\be
 S \ge S\ ;\quad
S \ge S'\ {\rm and}\  S'\ge S \Rightarrow S = S'\ ; \quad
S \ge S' \
\ee
and
\be
\ S' \ge S''\ \Rightarrow S
\ge
S''\ ;
\label{e2.1a}
\ee
and, given any $S', S'' \in L$, there exists $S \in L$ such that
\be
 S \ge S' \quad {\rm and} \quad S \ge S''\ .
\ee
$L$ will serve as the label set.  A {\it projective family}
$(\X_S,p_{SS'})_{S,S'\in L}$ consists of sets $\X_S$ indexed by
elements of $L$, together with a family of surjective {\it projections},
\be
p_{SS'}\:\ \X_{S'}\rightarrow \X_{S},
\ee
assigned uniquely to pairs $(S',S)$ whenever $S'\ge S$
such that
\be
p_{SS'}\circ p_{S'S''} = p_{SS''}   \ .
\ee
We will assume that $\X_S$ are all topological, compact, Hausdorff
spaces and that the projections $p_{SS'}$ are continuous.

In the application of this framework to theories of connections,
carried out in Sec. \ref{s4}, the labels $S$ can be thought of as
general lattices (which are not necessarily rectangular) and the
members $\X_{S}$ of the projective family, as the spaces of
configurations associated with these lattices.  The continuum theory
will be recovered in the limit as one considers lattices with
increasing number of loops of arbitrary complexity.

Note that, in the projective family there will, in general, be no set
${\Xb}$ which can be regarded as the largest, from which we can
project to any of the $\X_S$. However, such a set does emerge in an
appropriate limit, which we now define. The {\it projective limit}
$\Xb$ of a projective family $(\X_S, p_{SS'})_{SS'\in L}$ is the
subset of the Cartesian product $\times_{S\in L}\X_S$ that satisfies
certain consistency conditions:
\ba
\Xb\ :=\ \{(x_S)_{S\in L}& &\in
\times_{S\in L}\X_S\ : \nonumber\\
& &\ S'\ge S \Rightarrow p_{SS'}x_{S'} =
x_S\}.  \label{b2.4}
\ea
(This is the limit that  gave us in Sec. \ref{s4} the quantum configuration
$\agb$ for theories of connections.)
We provide $\Xb$ with the product topology that
descends from $\times_{S\in L}\X_S$. This is the {\it Tychonov
topology}.  In  the Tychonov  topology the product space is known to be
compact
and Hausdorff.  Furthermore, as noted in \cite{MM}, $\Xb$ is closed in
$\times_{S\in L} \X_S$, whence $\Xb$ is also compact (and
Hausdorff).
Note that the limit $\Xb$ is naturally
equipped with a family of projections:
\be
 p_{S}\ : \ \Xb \rightarrow \X_{S}, \ \ p_{S} ((x_{S'})_{S'
\in L}):= x_{S}\ .
\ee
Next, we introduce certain function spaces. For each $S$ consider
the space $C^0(\X_S)$ of the complex valued, continuous functions on
$\X_S$. In the union
\be
\bigcup_{S\in L} C^0(\X_{S})
\ee
we define the following equivalence relation. Given $f_{S_i}\in
C^0(\X_{S_i})$, $i=1,2$, let us say:
\be
f_{S_1} \ \sim\ f_{S_2}\ \ \ {\rm if}\ \
\ p_{S_1S_3}^* \ f_{S_1}\ =\ p_{S_2S_3}^*\ f_{S_2}
\label{b2.6}
\ee
for every $S_3\ \ge S_1, S_2$, where $p^*_{S_1,S_3}$ denotes the
pull-back map from the space of functions on $\X_{S_1}$ to the space
of functions on $\X_{S_3}$.
Using the equivalence relation we can now introduce the set of
{\it cylindrical functions} associated with the projective
family $(\X_S,p_{SS'})_{S,S'\in L}$,
\be
\Cyl(\Xb) \ := \ \big( \bigcup_{S\in L}C^0(\X_S)\ \big)
\ /\ \sim  \ .
\label{b2.9}
\ee
The quotient just gets rid of a redundancy: pull-backs of functions
from a smaller set to a larger set are now identified with the
functions on the smaller set.  Note that in spite of the notation, as
defined, an element of $\Cyl(\Xb)$ is {\it not} a function on ${\Xb}$;
it is simply an equivalence class of continuous functions on some of
the members $\X_S$ of the projective family. The notation is, however,
justified because, one {\it can} identify elements of $\Cyl(\Xb)$ with
continuous functions on $\Xb$.  This identification was implicitly
used in (\ref{e427}).  If the $\X_S$ are differentiable manifolds then
one can define spaces $\Cyl^n(\Xb)$ of differentiable cylindrical
functions in a a completely analogous way. These spaces play a crucial
role in defining measures and regulated operators.

\end{section}

\begin{section} {Unexpected consequences of Diffeomorphism invariance}
\label{unex}

In section \ref{phy}, we used a group averaging procedure to solve the
quantum diffeomorphism constraint. It was therefore natural to use the
finite --rather than the infinitesimal-- form of constraints. It turns
out, however, that there is really no choice: it is not possible to
define the infinitesimal form of the diffeomorphism constraints on
{\it any} $\Ha = L^2(\agb,d\mu)$ which carries a faithful
representation of the holonomy algebra when $d\mu$ is diffeomorphism
invariant. In this appendix, we will discuss this somewhat surprising
technical point.

Let $N^a$ denote a complete analytic vector field on $\S$ and
$\varphi_t$ the corresponding flow of analytic
diffeomorphisms. Then, from (\ref{2.2}), we see that the smeared
version of the diffeomorphism constraint is given by:
\ba
V_{N} &:=& \int_\S N^a(x) V_a(x) d^dx \nonumber\\
&=& \int_\S N^a(x) {\rm tr}\; [F_{ab}(x)\wt E^b(x)] d^dx \ .
\label{ddcc}
\ea
Let us equip $\ag$ with one of the standard Sobolev topologies
\cite{MV} and denote by $\widetilde N$ and $\widetilde \varphi_t$ the
vector field and the flow on $\ag$ induced by $N^a$. Given a smooth
function $\psi$ on $\ag$, it is then easy to write out the action of
the desired operator $\hat{V}_{N}$ on $\psi$:
\be \hat{V}_{N}\circ\psi
= \int_\Sigma d^dx\; N^a {\rm tr} F_{ab} {\delta \psi\over
\delta{A_b}}\ = {\cal L}_{\tilde N} \psi\ .
\ee
Hence, the exponentiated version of the constraint is given simply by:
$U_{N}(t)\circ \psi = (\tilde\varphi_t)_\star\cdot\psi$. Since
$\tilde\varphi$ extends naturally to $\agb$, it is straightforward to
extend the action of $U_{N}$ to our $\Ha$, which we will denote again
by $U_{N}$. If the measure on $\agb$ is diffeomorphism invariant,
$U_{N}$ are unitary operators, hence defined on all of $\Ha$. It is
now obvious that the algebra of these operators is closed; {\it there
are no anomalies}. The result is, however, non-trivial because our
constraint operators $U_N(t)$ are rigorously defined on the auxiliary
Hilbert space. It is known, for example, that if one uses a lattice
regularization to give meaning to the formally defined constraint
operators, anomalies do result.

What would happen if we try to extend to $\Ha$ the action of the
infinitesimal constraints $\hat{V}_N$ instead? Since Wilson loop
variables are smooth functions on $\ag$, let us begin by
setting $\psi(A) = T_\alpha(A)$ on $\ag$. Then, we have:
\be
\left(\hat{V}_N\circ  T_\a \right) (A) = \lim_{t \rightarrow 0}
{T_{\a_t} - T_\a \over t }(A) \ ,
\ee
where $\a_t = \phi_t \a$ and the point $A$ indicates that the limit is
taken pointwise in $\ag$. The limit is of course a well-behaved smooth
function on $\ag$. However, it fails to be a cylindrical function.
(Note that $U_N(t)\circ T_\alpha = T_{\phi(t)\cdot\alpha}$, on the
other hand, {\it is} cylindrical.) Hence, one might suspect that there
may be a difficulty in extending the operator $\hat{V}_N$ to $\Ha$. We
will see that this is the case.

More precisely, we now show that for a diffeomorphism invariant
measure $\mu$ on $\agb$ to be compatible with a well defined
infinitesimal generator of the diffeomorphism constraint, $\mu$ must
have a very special support.  The resulting representation of the
$\hat{T}_{\alpha}$ algebra would then be so unfaithful as to be
physically irrelevant.

Indeed, let $\mu$ denote a diffeomorphism invariant measure and $\a_t
= \varphi_t \a$ as above.  For the diffeomorphism constraint to be
well defined we must have (at least for ``most'' of the loops $\a$ in
$\S$)
\be
\lim_{t \rightarrow 0} \parallel T_{\a_t} -  T_\a \parallel^2_{L^2}
= \lim_{t \rightarrow 0} \int_{\agb}  (T_{\a_t} - T_\a)^2 d \mu = 0    \  .
\label{3.10}
\ee
{}From diffeomorphism invariance of the measure it is clear that
$$
\int_{\agb} T_\a^2 d\mu = \int_{\agb} T_{\a_t}^2 d \mu \ , \qquad \forall t
$$
and that there exists $ t_0 > 0$ such that
\be
\label{D6}
\int_{\agb} T_\a T_{\a_t} d \mu = k = const  \ , \qquad for \ t: \ 0 < t <
t_0 \ .
\ee
(To see this we can consider a flow $\varphi'_s$ of analytic
diffeomorphisms that leave $\a$ invariant and such that $\varphi'_s
\a_t = \a_{ t'(t,s)}$.)
For the limit in (\ref{3.10}) to be equal to zero we must have $k =
\int_{\agb} T_\a^2 d \mu$, which from (\ref{3.10})  implies that in
$L^2(\agb, \mu)$
\be
T_{\a_t} = T_\a      \ , \qquad \forall t \ : \ 0 < t < t_0   \   .
\label{nfac}
\ee
Now, (\ref{nfac}) implies that the representation $\rho$ (see
(\ref{3.9})) of the holonomy algebra on $L^2(\agb, \mu)$ is not
faithful since $T_{\a_t} - T_\a \neq 0$ as elements of $\cal HA$,
while $\rho ( T_{\a_t} - T_\a ) = 0$ as operators on $L^2(\agb, \mu)$.
Thus, the support of the measure $\mu$ is so special that it is not
suitable as a kinematical measure in quantum theory. Put differently,
in any interesting representation of the holonomy algebra,
\be
\int_{\agb} T_{\a_t} T_\a d\mu \neq \int_{\agb} T_\a^2 d\mu \ , \qquad
\forall t \ : \ 0 < t < t_0    \  \label{3.11}
\ee
and therefore the infinitesimal generators of the diffeomorphism
constraints can not be well defined.

\end{section}

\begin{section} {Geometrical Operators}
\label{go}

On the phase space of Riemannian general relativity, the momentum
variable $\tilde{E}^a_i$ has the interpretation of a density weighted
triad. Hence, one can use it to construct functions on the phase space
that carry geometrical information. For example, the volume of a
region $R$ within $\Sigma$ is be given by:
\be
V_R := \int_R d^3x |\eta_{abc} \epsilon^{ijk} \tilde{E}^a_i
\tilde{E}^b_j \tilde{E}^c_k|^{1\over 2} \ ,
\ee
where $\eta_{abc}$ is the Levi-Civita tensor density on
$\Sigma$.  Similarly, the area of a 2-surface $S$ within $\Sigma$
defined by, say, $x_3 ={\rm const}$ is given by:
\be
A_S := \int_S d^2x |\tilde{E}^a_i \tilde{E}^{bi} \nabla_a x_3
\nabla_b x_3 |^{1\over 2}\,
\ee
The question then arises: are there well-defined geometric operators
$\hat{V}_R$ and $\hat{A}_S$ on $\Ha$? In absence of matter fields,
$V_R$ and $A_S$ fail to be observables since they are not
diffeomorphism invariant. Hence, the corresponding operators will not
represent physical observables. However, if we bring in matter sources
and {\it define} the regions $R$ and surfaces $S$ using these fields,
{\it then} $\hat{V}_R$ and $\hat{A}_S$ would be observables with
respect to the diffeomorphism constraints \cite{matter}. Therefore, it
is of considerable interest to try to construct these operators in the
kinematical setting of Sec.IV and explore their properties.

At first sight, it seems difficult to make sense out of these
operators.  To begin with, $\hat{E}^a_i$ itself is not a well-defined
operator on $\Ha$. Second, the desired operators would require {\it
products} of $\hat{E}^a_i$ evaluated at the same point, and,
furthermore, a square-root! Nonetheless, it turns out that these
formal expressions {\it can be} regulated satisfactorily to yield
well-defined operators on $\Ha$. The regularization procedure involves
point-splitting and it is necessary to fix a gauge and a background
metric (or coordinate system) in the intermediate stage. However, when
the regulator is removed, the final expression is not only
well-defined but independent of the background structures used in the
procedure. The overall procedure is similar to the one used in
rigorous quantum field theories. Furthermore, somewhat surprisingly,
for suitable operators such as $\hat{V}_R$ and $\hat{A}_S$, the
situation is better than what one might have expected: there is no
need to renormalize, whence the final answers have no free
parameters. Finally, the operators are essentially self-adjoint on
$\Ha$ and their spectra are often discrete.  Thus, the ``quantum
geometry'' that emerges from our framework has certain essentially
discrete elements which suggest that the use of a continuum picture at
the Planck scale is flawed.  These results are analogous to the ones
obtained by Rovelli and Smolin \cite{24} in the loop
representation. However, the precise relation is not known.

Here, we will illustrate these results with the area operator. For
simplicity, let us suppose that we can choose coordinates on $\Sigma$
in a neighborhood of $S$ such that $S$ is given by $x_3 = {\rm const}$
and $x_1, x_2 $ coordinatize $S$.  Then, we can write $A_S$ as: $A_S =
\int_S d^2x\ \sqrt|O(x)|$, where $O(x) = \tilde{E}^3_i(x)
\tilde{E}^{3i}(x)$. To define $\hat{A}_{S}$, let us use a point splitting
procedure and consider the regulated operator
\be \label{5.3.10}
\wh{O}_\epsilon (x) := \int d^3y f_\epsilon(x,y)\int d^3z
f_\epsilon(x,z) \frac{\delta}{\delta A_3^i(y)} \frac{\delta}{\delta
A_3^i(z)}\; ,
\ee
where $f_\epsilon(x,y)$ (is a density of weight $1$ in $x$ and function
in $y$ and that) tends to $\delta^3(x,y)$ in the limit. For concreteness,
we will construct it from $\Theta$ density/functions:
\ba
f_\epsilon(x,y) = {1\over{\epsilon^3}} [\Theta
(\textstyle{\epsilon\over 2}& & - |x_1 -y_1|) \Theta
(\textstyle{\epsilon\over 2} - |x_2 -y_2|) \times  \nonumber\\
& &\Theta(\textstyle{\epsilon\over 2} - |x_3 -y_3|)] \  .
\ea
(There is thus an implicit background density of weight one in $x$ in
the expression of $\Theta$.)

Now, let us begin by considering a cylindrical function $F_\gamma$ on
the space $\A$ of smooth connections. By using a group-valued chart on
$\A_\gamma$, $F_\gamma$ can be expressed as $F_\gamma (A) = f(g_1,
..., g_N)$ where $N$ is the number of edges in $\gamma$ and $g_I =
{\cal P} \exp \int_{e_I} A $. A simple calculation yields:
\ba \label{5.3.11}
& & \wh{O}_\epsilon (x) \circ  F_\gamma\nonumber\\
&=&\int d^3y f_\epsilon(x,y)\
\frac{\delta}{\delta A_3^i(y)}
\sum_{e_J} \int_{e_J} ds\; f_\epsilon(x,e_j(s))\nonumber \\
& &\dot{e}^3_J\;  \mbox{Tr}(h_J(1,s)\tau_i h_J(s,0)
\frac{\partial}{\partial h_J}) f_\gamma\nonumber\\
&=& \sum_{I,J} \int_{e_I} dt\  \dot{e}^3_I(t) \int_{e_J} ds\
\dot{e}^3_J(s)\ f_\epsilon(x,e_I(t))\times  \nonumber\\
& &f_\epsilon(x,e_J(s)) \mbox{Tr}(h_I(1,t)\tau_i h_I(t,0)\frac{\partial}
{\partial h_I})\times\nonumber\\
& & \mbox{Tr}(h_J(1,s)\tau_i h_J(s,0)\frac{\partial}{\partial h_J}) f_\gamma
\nonumber\\
& & +2\sum_I \int_{e_I\times e_I,t\ge s}dt ds\  \dot{e}^3_I(t)\
\dot{e}^3_J(s)\nonumber\\
& & f_\epsilon(x,e_I(t)) f_\epsilon(x,e_J(s))
\nonumber\\
& & \big[\;\mbox{Tr}(h_I(1,t)\tau_i h_I(t,s)\tau_j
h_I(s,0)\frac{\partial}{\partial h_I})+\nonumber\\
& &\mbox{Tr}(h_I(1,t)\tau_i h_I(t,s)\tau_j h_I(s,0)\frac{\partial}
{\partial h_I})\big] f_\gamma \nonumber\\
& &=: (\hat{O}_\epsilon^I(x) + \hat{O}_\epsilon^{II}(x))\circ f
\ea
The right side is a well-defined function of smooth connections $A$.
However, it is no more a cylindrical function because of the form of
the terms involving integrals over edges. We thus have two problems:
the action of $O_\epsilon (x)$ is not well-defined on functions of
{\it generalized} connections, and, even while operating on functions
of smooth connections, the operator sends cylindrical functions to
more general ones. We will see that the two problems go away once the
regulator is removed.

Let us consider the first term in detail; an analogous treatment of
the second term shows that it does not contribute to the final result.

Ultimately, we want to integrate $\hat O(x)$ over $S$. Hence, we want
$x$ to lie in $S$. Then, for sufficiently small $\epsilon$, because of
the $f_\epsilon$ terms, only the edges that intersect $S$ contribute
to the sum. (Furthermore, since only the third component of the
tangent vectors count in $O_\epsilon^I$, edges which lie within $S$ do
not contribute.)  Without loss of generality, we can assume that
intersections occur only at vertices of $\gamma$ (since we can always
add vertices in the beginning of the calculation to ensure this). Now,
if we write out the functions $f_\epsilon$ explicitly and Taylor
expand, around each vertex at which $\gamma$ intersects $S$, the group
elements that appear in the integrals we can express
$O_\epsilon^I\circ f$ as a sum:
\ba
& &\hat{O}_\epsilon^I(x)\circ f =
\sum_{v_\alpha}\sum_{I_\alpha}\sum_{J_\alpha}\;
{K(I_\alpha, J_\alpha)\over \epsilon^4}\nonumber\\
& & [ \Theta({\epsilon\over 2}
- |x_1-x_1^{(\alpha)}|)
\Theta({\epsilon\over 2}-|y_1-y_1^{(\alpha)}|)]^2\times\nonumber\\
& & \{X_{I_\alpha,i}X_{J_\alpha,i} + o(\epsilon)\}
\circ f (g_1, ...,g_n)\; .
\ea
Here $v_\alpha$ are the vertices of $\gamma$ that lie in $S$,
$e_{I_\alpha}, e_{J_\alpha}$ are the edges passing through the vertex
$v_\alpha$, $X_{I_\alpha}$ is the right (left) invariant vector field
on the copy of the group corresponding to the edge $e_{I_\alpha}$
which points at the identity of the group in the $i$-th direction, if
the edge is oriented to be outgoing (incoming) at the vertex, and the
constant $K(I_\alpha, J_\alpha)$ equals $+1$ if the two edges lie on
the opposite side of $S$, $-1$ if they lie on the same side and
vanishes if the tangent vector of either edge is tangential to $S$.

Let us try to take the limit of $\hat{O}_\epsilon^I(x) \circ f$ as
$\epsilon$ tends to zero. In this limit, each $\Theta_\epsilon$ tends
to a 1-dimensional Dirac $\delta$-distribution, and the expression
then diverges as $1\epsilon^2$. As is usual in field theory, we can
first renormalize the expression by $\epsilon^2$ and {\it then} take
the limit. Now, the limit clearly exists. However, it depends on the
background density implicit in the expression of $\Theta$ and hence
the resulting operator carries the memory of the background structure
used in the regularization. That is, the ambiguity in the final answer
is not of a multiplicative constant, but of a background density of
weight one. (This is to be expected since the left hand side is a
density of weight $2$ (in $x$ and $y$) while the 2-dimensional Dirac
$\delta$-distribution is only a density of weight $1$.) Because of the
background dependence, the resulting operator is not useful for our
purposes.

However, if we take the square-root of the regulated operator and
{\it then} take the limit, we obtain a well-defined result:
\ba
& & \lim_{\epsilon\to 0} {|\hat O_\epsilon(x)|}^{1\over 2}
\circ f =\sum_{v_\alpha} \delta^2(x, v_\alpha)\times\nonumber\\
& & |\sum_{I_\alpha}\sum_{J_\alpha}\;
(K(I_\alpha, J_\alpha)\{X_{I_\alpha,i}X_{J_\alpha,i}\}|^{1\over 2}
\circ f (g_1, ...,g_n)\; .
\ea
Note that, now, no renormalization is necessary. In the final result,
both sides are densities of weight one and there is neither background
dependence, nor any free parameters.  With proper specification of
domains, the operator under the square-root can be shown to be a
non-negative self-adjoint operator on $L^2((SU(2))^n, d\mu_H)$. (For
example, if there are just two edges at a vertex $v_\alpha$, one on
each side of $S$, then the operator is just the (negative of the)
Laplacian.) Hence, the square-root is well-defined.  We can therefore
construct an area operator:
\be
\hat{A}_S \circ F_\gamma = \int_S d^2x\; {|\hat O(x)|}^{1\over 2}\circ
f\; .
\ee
Clearly, this operator maps cylindrical functions to cylindrical
functions. It is straightforward to show it satisfies the
compatibility conditions discussed in Sec. \ref{s4}  and thus
leads to a well-defined operator on $\Ha=L^2(\agb, d\mu_o)$.  This
operator is self-adjoint and has a discrete spectrum.

The volume operator can be treated in a similar manner.

To conclude, note that there is a striking qualitative resemblance
between this analysis of properties of geometry and that of physical
properties of polymers in condensed matter physics \cite{prigodin}.
In both cases, the basic excitations are ``loopy'' rather than
``wavy''; they reside along $1$-dimensional graphs rather than on
3-dimensional volumes. However, under suitably complex conditions,
they resemble genuinely 3-dimensional systems \cite{prigodin,IR}.

\end{section}


\begin{references}


\bibitem{1} V.\ Husain, K.\ Kucha\v{r}, Phys.\ Rev. {\bf D42}, 12 (1990).

\bibitem{2} A.\ Ashtekar, Phys.\ Rev. Lett.\ {\bf 57} 2244 (1986),
            Phys.\ Rev.\ {\bf D36}, 1587 (1987).

\bibitem{6} A.\ Ashtekar, J.\ Lewandowski, D.\ Marolf, J.\ Mour\~ao, T.\
            Thiemann, ``Coherent State Transform for Spaces of Connections",
            J. Funct. Analysis (in press), preprint gr-qc/9412014.

\bibitem{8} A.\ Ashtekar, J.\ D.\ Romano, Physics Letters {\bf B229}, 56
            (1989).

\bibitem{AI} A. Ashtekar and C.J. Isham,
Class. \& Quan. Grav. {\bf 9}, 1433 (1992).

\bibitem{AL1} A. Ashtekar and J. Lewandowski, ``Representation
theory of analytic holonomy $C^\star$ algebras'', in {\it Knots and
quantum gravity}, J. Baez (ed), (Oxford University Press, Oxford 1994).

\bibitem{B2} J. Baez, Lett. Math. Phys. {\bf 31}, 213 (1994);
``Diffeomorphism invariant generalized measures on the space of
connections modulo gauge transformations", hep-th/9305045,
in the Proceedings of the conference on quantum topology, D. Yetter
(ed) (World Scientific, Singapore, 1994).

\bibitem{L} J. Lewandowski, Int. J. Mod. Phys. {\bf D3}, 207 (1994).

\bibitem{MM} D. Marolf and J. M. Mour\~ao, ``On the support of the
Ashtekar-Lewandowski measure'',  Commun. Math. Phys.,
(in press).

\bibitem{AMM} A. Ashtekar, D. Marolf and J. Mour\~ao, ``Integration on
the space of connections modulo gauge transformations'' in {\it The
Proceedings of the Lanczos International Centenary Conference}, J. D.
Brown, et al (eds) (SIAM, Philadelphia, 1994).

Jerzy Lewandowski,  ``Differential Geometry for the Space of Connections
Modulo Gauge Transformations'',  in {\it The
Proceedings of the Lanczos International Centenary Conference}, J. D.
Brown, et al (eds) (SIAM, Philadelphia, 1994).

\bibitem{AL2} A. Ashtekar and J. Lewandowski, ``Differential
geometry on the space of connections via graphs and projective
limits'', preprint CGPG-94/12-4, hep-th/9412073.

\bibitem{AL3} A. Ashtekar and J. Lewandowski, J. Math. Phys. {\bf 36}, 2170
(1995).

\bibitem{AH2} A. Higuchi Class. Quant. Grav. {\bf 8},  1983 (1991).

\bibitem{AH3} A. Higuchi Class. Quant. Grav. {\bf 8}, 2023 (1991).

\bibitem{DD} D. Marolf, ``The spectral analysis inner product for
quantum gravity,'' preprint gr-qc/9409036, to appear in the
Proceedings of the VIIth Marcel-Grossman Conference, R. Ruffini and
M. Keiser (eds) (World Scientific, Singapore, 1995); D. Marolf,
Ph.D. Dissertation, The University of Texas at Austin (1992).

\bibitem{DM1} D. Marolf, ``Quantum observable and recollapsing
dynamics,'' preprint gr-qc/9404053. Class. Quant. Grav. (1995) (in
press).

\bibitem{A2} A. Ashtekar, {\it Non-Perturbative Canonical Gravity},
Lectures notes prepared in collaboration with R.S. Tate (World Scientific,
Singapore, 1991); in {\it Gravitation and Quantization}, B. Julia (ed)
(Elsevier, Amsterdam, 1995).

\bibitem{AT} A. Ashtekar and R. S. Tate,  J. Math. Phys.{\bf 35}, 6434
             (1994).

\bibitem{29} P.\ Hajicek, in : Proceedings, Bad Honnef, Germany, 1993,
            J.\ Ehlers, H.\ Friedrich (eds.), Lecture notes in Physics,
            (Springer-Verlag, Berlin, New York, 1994).

\bibitem{24a} J.\ Baez, ``Spin network states in gauge theory",
Adv. Math. (in press); ``Spin networks in non-perturbative quantum
gravity,'' pre-print gr-qc/9504036.

\bibitem{24} C.\ Rovelli, L.\ Smolin, ``Discreteness of volume and
area in quantum gravity'' Nucl. Phys. B (in press); ``Spin networks and
quantum gravity'' pre-print CGPG-95/4-4.

\bibitem{23} T.\ Thiemann, ``The inverse Loop Transform", preprint
CGPG-94/4-1.

\bibitem{G} R. Gambini, A. Trias, Nucl. Phys. {\bf B278}, 436 (1986).

\bibitem{RS} C. Rovelli, L. Smolin, Nucl. Phys. {\bf B331}, 80 (1990).

\bibitem{S} R. Rovelli, Class. Quan. Grav. {\bf 8}, 1613 (1991);
L. Smolin, in {\it Quantum Gravity and Cosmology}, J.  P\'erez-Mercader
et al eds (World Scientific, Singapore, 1992); R. Gambini, in the {\it
Proceedings of the IVth Mexican Workshop on particles and Fields}
(World Scientific, Singapore, in press).

\bibitem{TT} T.\ Thiemann, Class. Quantum Grav.\ {\bf 12}, 59 (1995).

\bibitem{22} I.\ M.\ Gel'fand, N.\ Ya.\ Vilenkin, ``Generalized Functions",
            vol. 4, Applications of Harmonic Analysis, Academic Press,
            New York, London, 1964.

\bibitem{RS1} M Reed and B Simon 1970 {\it Functional Analysis,
Meth. Mod. Math. Phys. Vol I} (New York: Academic Press).

\bibitem{DM3}
D. Marolf ``Almost Ideal Clocks in Quantum Cosmology: A Brief
Derivation of Time,'' preprint gr-qc/9412016.

\bibitem{5} A.\ Ashtekar, J.\ Lewandowski, D.\ Marolf, J.\ Mour\~ao, T.\
            Thiemann, ``Quantum holomorphic connection dynamics", in
            preparation.

\bibitem{9} J. D. Romano, Gen. Rel. \& Grav. {\bf 25}, 759 (1993).

A.\ Ashtekar, V.\ Husain, C.\ Rovelli, J.\ Samuel, L.\ Smolin,
            Class. Quantum Grav.\ {\bf 6},  L183 (1989).

\bibitem{GJ} J. Glimm and A. Jaffe, ``Quantum physics",
(Springer-Verlag, New York, 1987).

\bibitem{R} V. Rivasseau, ``From perturbative to constructive
renormalization", (Princeton University Press, Princeton, 1991).

\bibitem{AM} M. Asorey and P.K. Mitter,
Commun.  Math. Phys. {\bf 80}, 43 (1981); M. Asorey and F. Falceto,
Nucl Phys. {\bf B327}, 427 (1989).

S. Klimek, W. Kondracki, Commun. Math. Phys.{\bf 113}, 389 (1987).

\bibitem{Ba} J.W. Barret, Int. Journ. Theor. Phys.
{\bf 30} 1171,  (1991).

J. Lewandowski, Class. Quant. Grav. {\bf 10}, 879 (1993).

A. Caetano, R. F. Picken, ``An axiomatic definition of
holonomy'', Int. Jour. Math. {\bf 5}, 835 (1994).

\bibitem{jbaa} J. Baez, private communication to A. Ashtekar, 1994.

\bibitem{ALMMT1} A. Ashtekar, J. Lewandowski, D. Marolf, J. Mour\~ao
and T. Thiemann, ``A manifestly gauge invariant approach to quantum
theories of gauge fields'', in {\it Geometry of constrained dynamical
systems}, J. Charap (ed) (Cambridge University Press, Cambridge,
1994); ``Constructive quantum gauge field theory in two space-time
dimensions'' (CGPG preprint).

\bibitem{Ya} Y. Yamasaki, ``Measures on infinite dimensional spaces,''
(World Scientific, Singapore, 1985).

\bibitem{DWM} C. DeWitt-Morette, Commun. Math. Phys. {\bf 28}, 47 (1972);
Commun. Math. Phys. {\bf 37} 63 (1973).

\bibitem{MV} P. K. Mitter and P. Viallet,  Comm. Math. Phys.
{\bf 79}, 43 (1981).

\bibitem{matter} C. Rovelli, Nucl. Phys. {\bf B405}, 797 (1993);
L. Smolin, Phys. Review {\bf D49}, 4028 (1994).

\bibitem{prigodin} V. N. Prigodin, K. B. Efetov, Phys. Review Lett.
{\bf 70}, 2932 (1993); Class. Quantum. Grav.

\bibitem{IR} J. Iwasaki and C. Rovelli, Int.  J. Modern. Phys.
{\bf D1}, 533 (1993); Class. Quant. Grav. {\bf 11}, 2899 (1994).

\end{references}
\end{document}